\documentclass[a4paper,11pt]{article}
\pdfoutput=1
\usepackage{jheppub}
\usepackage[T1]{fontenc}
\usepackage{amsmath,amssymb,mathtools,bm,mathrsfs}
\usepackage{booktabs}
\usepackage{array}
\usepackage{tikz}
\usetikzlibrary{arrows.meta}

\makeatletter
\gdef\@fpheader{}
\makeatother

\newcommand{\ii}{\mathrm{i}}
\newcommand{\dd}{\mathrm{d}}
\newcommand{\Del}{\Delta_0}
\newcommand{\Sig}{\Sigma}

\newcommand{\cK}{\mathsf{K}}

\newcommand{\cM}{\mathcal{M}}
\newcommand{\cE}{\mathcal{E}}
\newcommand{\cU}{\mathcal{U}}
\newcommand{\cF}{\mathcal{F}}

\title{\boldmath On scalar and electromagnetic perturbations of the root--Kerr object}

\author[a,b]{Samim Akhtar}
\affiliation[a]{ICTP South American Institute for Fundamental Research,\\
Rua Dr. Bento Teobaldo Ferraz 271-2, 01140-070, S\~ao Paulo, SP, Brazil}
\affiliation[b]{Instituto de F\'isica Te\'orica UNESP -- Universidade Estadual Paulista,\\
Rua Dr. Bento Teobaldo Ferraz 271, 01140-070, S\~ao Paulo, SP, Brazil}
\emailAdd{samim.akhtar@ictp-saifr.org}

\abstract{%
We study scalar and electromagnetic perturbations of the root--Kerr object.  Its
source-free exterior carries the electromagnetic field obtained in
the vanishing-mass limit of the Kerr--Newman black hole, while its physical
source is a rotating disk with an essential distributional rim contribution.
For a massless charged scalar, the minimally coupled Klein--Gordon equation
separates into a confluent-Heun (CHE) radial equation with a direct coupling
between the scalar and background charges.  We give the associated
Nekrasov--Shatashvili (NS) dictionary and show that the disk is an ordinary
radial point.  For electromagnetic pertubations, we derive the equations for the radiative
Newman--Penrose Maxwell scalars directly.  They have the same confluent-Heun
structure, but a Maxwell perturbation has no bulk coupling to the
background potential.  Charge-dependent photon scattering data therefore
cannot be fixed by the exterior equation alone.  We formulate the disk/rim
boundary-data problem, identify the response data required to convert
exact CHE/NS connection coefficients into a physical scattering matrix, and
show how all-spin helicity-flip Compton information constrains the response
on the radiative subspace.  The analysis separates exact exterior propagation
and amplitude matching from the additional source dynamics required
for a microscopic response.}

\begin{document}
\maketitle
\flushbottom

\section{Introduction and summary}
\label{sec:intro}

The scattering of waves by spinning compact objects lies at the intersection
of black-hole perturbation theory, classical effective field theory, and
on-shell amplitudes.  In the perturbative description, one separates the
field equations in the geometry of a Kerr black hole, imposes a condition at the future horizon,
and extracts observables from the radial connection problem.  The
Newman--Penrose formalism and the Teukolsky master equation furnish the
standard framework for this construction \cite{Newman:1961qr,Teukolsky:1973ha}.
The relation of black-hole scattering to massive-particle Compton amplitudes
has subsequently been established in a sequence of calculations that combine
long-wavelength perturbation theory, partial waves, and higher-spin
amplitudes \cite{Bautista:2021wfy,Bautista:2022wjf,Saketh:2022wap,
Bautista:2023sdf}.

Independently, amplitude methods have exposed a highly constrained all-spin
structure associated with the multipoles of Kerr black holes.  Minimal coupling, coherent-spin
representations, and classical limits provide complementary descriptions of
this structure \cite{Arkani-Hamed:2017jhn,Chung:2018kqs,Guevara:2017csg,
Arkani-Hamed:2019ymq,Chung:2019yfs,Guevara:2019fsj,Maybee:2019jus,
Vines:2018gqi,Aoude:2020onz,Aoude:2021oqj,Alessio:2023kgf}.  These developments have motivated
systematic analyses of classical Compton amplitudes, spinning scattering,
and their worldline realizations \cite{Chiodaroli:2021eug,Aoude:2022trd,
Aoude:2022thd,Chen:2022clh,Haddad:2023ylx,
Aoude:2023vdk,Bjerrum-Bohr:2023jau,Bjerrum-Bohr:2023iey,
Cangemi:2022bew}.  The programme is particularly valuable because it
separates universal on-shell information from finite-size response data, a
distinction that is also central in effective descriptions of compact-object
scattering \cite{Saketh:2022wap,Akhtar:2024mbg}.

The connection between these amplitude constructions and the classical
double-copy programme provides a further motivation for the present problem.
In the massive higher-spin framework, the root--Kerr solution is the
gauge-theory counterpart of the Kerr multipole structure: its electromagnetic
three- and four-point couplings organize data that are closely related to the
corresponding Kerr amplitudes \cite{Bern:2019crd,Haddad:2020tvs,
Alessio:2023kgf,Cangemi:2022bew,Cangemi:2023ysz}.  This relation does not
identify the microscopic source of the two solutions.  Rather, it makes the
root--Kerr configuration a controlled single-copy setting in which the all-spin
multipole tower is retained while the exterior geometry is flat and the
disk/rim source is explicit.

This setting is particularly valuable because higher-spin Compton amplitudes
are not fixed by their three-point couplings alone.  Naive minimal-coupling
factorizations develop unphysical spurious poles beyond spin one for photon
Compton scattering and beyond spin two for gravitational Compton scattering.
The local contact terms required to remove these singularities are constrained
by locality, Ward identities, and the classical limit, but they carry
information that is not contained in the three-point amplitude
\cite{Chiodaroli:2021eug,Cangemi:2022bew,Cangemi:2023ysz,
Cangemi:2023bpe}.  In the root--Kerr problem, the same contact data must be
compatible with the response of the material disk and rim; this is the point
at which an on-shell amplitude meets a physical scattering problem.

The aim of this work is to establish that link at the level of perturbations.
The exterior connection problem propagates specified gauge-invariant boundary
data (disk/rim traces) to asymptotic channels, while the root--Kerr Compton
amplitude constrains the local on-shell reaction in the same space of boundary
data.  A microscopic response law is required to combine these ingredients
into a physical scattering matrix.
For Kerr black holes, regularity on the future horizon selects the retarded,
purely ingoing solution.  This condition permits finite-size and absorptive
effects to be extracted from the corresponding connection problem
\cite{Teukolsky:1973ha,Aminov:2020yma,Bautista:2023sdf,
Jones:2023ugm,Bautista:2024emt}.  In the root--Kerr setting, a
microscopic disk/rim law must play this selecting role.  Once specified, it
would make it possible to determine the associated gauge-theory response
before relating it to nonlinear gravitational observables through the
classical double copy.  The detailed post-Minkowskian order of a given
effect is then observable- and completion-dependent.

At the level of classical observables, post-Minkowskian effective field
theory, eikonal methods, and modern amplitude constructions provide complementary
treatments of conservative, radiative, and spin-dependent effects
\cite{Kalin:2020mvi,Liu:2021zxr,Kalin:2022hph,Manohar:2022dea,
Kalin:2019rwq,Kalin:2019inp,Cho:2021arx,Akhtar:2024soft,
AkhtarLaddhaMannaManu:2026,Foffa:2013qca,Foffa:2019eeb,
Foffa:2021pkg,Caron-Huot:2025tlq,Porto:2016pyg}.  These developments sharpen the need to
distinguish data intrinsic to a prescribed exterior field from information
associated with the dynamics of the compact source.

The radial equations that arise in black-hole perturbation theory are commonly
organized by the singularity structure of Heun equations and their confluences
\cite{Suzuki:1998vy,Hortacsu:2011rr}.  For the Kerr black hole, the connection
problem admits a nonperturbative description in terms of
Nekrasov--Shatashvili (NS) functions.  This description follows from the
Alday--Gaiotto--Tachikawa (AGT) relation between Virasoro conformal blocks and four-dimensional
supersymmetric gauge theory, together with the NS limit of the corresponding
instanton partition function \cite{Alday:2009aq,
Nekrasov:2009rc}.  Explicit connection formulae for Heun equations
and their confluences have been derived from irregular Liouville correlators
\cite{Bonelli:2022ten}.  Applications to black-hole
spectra, greybody factors, tidal response, and Compton amplitudes of Kerr black holes have
been developed in \cite{Aminov:2020yma,
Bonelli:2021uvf,Bautista:2023sdf,Akhtar:2025nmt}.  The resulting formulation organizes the
analytic-continuation data around singular points (monodromies), connection
coefficients, and controlled low-frequency expansions of the confluent-Heun
(CHE) equation.  It thereby provides a precise method for
problems sharing the radial singularity structure of the Kerr black hole problem but possessing a distinct
physical inner locus.  The root--Kerr object, whose exterior carries the
Lynden--Bell magic electromagnetic field \cite{Lynden-Bell:2002dvr}, provides such a problem.

The Newman--Janis construction produces a stationary, axisymmetric
electromagnetic configuration from a Coulomb seed, with multipoles generated by
a single length $a$ \cite{Newman:1965tw,Erbin:2016lzq}.  This configuration
is the $G\!M\to0$ limit of the Kerr--Newman black hole with charge and ring radius held
fixed.  Its metric is flat away from the source, whereas its Maxwell field is
nontrivial.  Lynden--Bell demonstrated that the physical source is
distributional: it consists of a rotating disk together with an essential rim
contribution \cite{Lynden-Bell:2002dvr}.  The same field has appeared as the
electromagnetic root of all-spin Compton amplitudes and higher-spin gauge
constructions \cite{Cangemi:2023ysz,Cangemi:2023bpe,Scheopner:2023rzp}.
It therefore permits a controlled distinction between
exact exterior propagation, on-shell scattering data, and a microscopic
linear-response law for the source.

That distinction is decisive here.  Root--Kerr has no horizon, and its real
source is distributional and supported on a disk and its rim.  A condition imposed on this source is
therefore not equivalent to a future-horizon ingoing condition at either
complex zero of the radial coefficient.  The confluent-Heun connection
problem determines exterior propagation between specified bases, whereas a
physical scattering matrix additionally requires a retarded disk/rim response
whenever the probe interacts with the source.  Conversely, a three- or
four-point amplitude constrains selected on-shell matrix elements; it does not
by itself determine an off-shell conductivity, a spectral density, or a local
surface equation of state.  This distinction is consistent with the general
effective-field-theory separation of conservative multipoles from dissipative
response \cite{Saketh:2022wap,Ivanov:2022hlo}.

The electromagnetic perturbation problem introduces a further structural issue.  Although the
root--Kerr object carries charge $Q$ through its magic exterior field and its Compton amplitudes contain
charge-dependent terms, a Maxwell perturbation does not minimally
couple to a prescribed Abelian background potential.  Its source-free
exterior spin-$1$ equation is consequently independent of $Q$.  The physical
$Q^2$ dependence of photon scattering must arise when the incident wave
induces disk/rim degrees of freedom.  This differs fundamentally from the
charged-scalar problem, in which minimal coupling to $\bar A_\mu$ enters the
radial operator directly.  Maintaining this distinction is necessary for a consistent
Newman--Penrose analysis and for a comparison with on-shell amplitudes.

Our results are as follows.
\begin{enumerate}
\item We derive the massless charged-scalar equation directly from minimal
coupling to the magic potential.  The separated radial equation contains the
product of the scalar charge and the background charge $Q$, is confluent Heun,
and provides a controlled scalar Nekrasov--Shatashvili connection problem.
The neutral scalar
is recovered when the scalar charge is set to zero.
\item We derive the neutral spin-$1$ exterior equation directly from the
linearized Newman--Penrose Maxwell system.  The background charge $Q$ drops
out of this equation.  Accordingly, the neutral spin-$1$ equation must be
derived from the Maxwell system; it does not follow from applying the
Newman--Janis construction to the charged-scalar Coulomb equation.
\item Both separated radial equations are confluent Heun equations with regular
singular points at $r=\pm\ii a$ and an irregular singularity at infinity.  We
give their NS connection data and identify the physical disk with the ordinary
point $z_D=1/2$.
\item We identify the all-spin root--Kerr Compton candidate of
ref.~\cite{Cangemi:2023ysz} as information about the response probed by
physical radiative photon modes.  Once boundary-data normalization and a
regulator are chosen, it fixes an unambiguous on-shell reaction on those
modes.  It does not fix the remaining disk/rim degrees of freedom and hence
does not constitute a microscopic material model.
\item We derive the distributional disk junction conditions and show that a
simple finite-speed regulator with one subluminal corotating rim charge cannot
reproduce both the total charge and the magnetic moment.  A physical
completion therefore requires an additional rim current or magnetization
channel, beyond the regulated disk matter.
\end{enumerate}

The manuscript is organized to make the logical chain explicit.  Section
\ref{sec:background} reviews the root--Kerr geometry and the physical disk
source.  Section \ref{sec:scalar} derives the charged-scalar problem and its
confluent-Heun and Nekrasov--Shatashvili data.  Sections \ref{sec:photon} and \ref{sec:ns} give the independent
Maxwell derivation, the two connection problems, and the physical matching
locus.  Section \ref{sec:response} explains the limited, on-shell role of
the amplitude data.  We close with the disk/rim constraint and the remaining
physical questions.  We treat the exterior connection problem using the NS
formalism.  Its connection matrix does not, however, determine a root--Kerr
phase shift until a disk/rim response condition has been specified.

\section{The root--Kerr object: exterior magic field and physical source}
\label{sec:background}

The term \emph{root--Kerr object} is used here in the following precise sense.
We start with the Kerr--Newman black hole in Boyer--Lindquist
coordinates and take the $G\!M\to0$ limit, with $G$ Newton's constant, while retaining the
electromagnetic charge $Q$ and the rotation parameter $a$.  Equivalently,
one may set the mass parameter to zero after the oblate coordinate system has
been selected.  The metric then becomes Minkowski space, but not in Cartesian
coordinates.  The vector potential remains nontrivial.  This limit is neither
a black hole nor a superextremal continuation with a horizon removed; nor is
it merely the field of a point charge.  The root--Kerr object comprises this
flat-space Maxwell exterior together with its distributional disk/rim source.
The terminology root--Kerr emphasizes the role of this configuration
as the electromagnetic seed of the multipole structure of Kerr black holes
\cite{Lynden-Bell:2002dvr,Cangemi:2023ysz}.

We put
\begin{equation}
 \Del=r^2+a^2,\qquad \Sig=r^2+a^2\cos^2\theta .
\end{equation}
The root--Kerr metric and magic potential are
\begin{align}
 \dd s^2&=-\dd t^2+\frac{\Sig}{\Del}\dd r^2+\Sig\dd\theta^2
       +\Del\sin^2\theta\,\dd\phi^2, \label{eq:metric}\\
 \bar A&=-\frac{Qr}{\Sig}\left(\dd t-a\sin^2\theta\,\dd\phi\right).
 \label{eq:potential}
\end{align}
The flatness of \eqref{eq:metric} can be seen directly from the Cartesian
map
\begin{equation}
 x+\ii y=(r+\ii a)e^{\ii\phi}\sin\theta,
 \qquad z=r\cos\theta .
 \label{eq:cartesianmap}
\end{equation}
In particular,
\begin{equation}
 x^2+y^2=(r^2+a^2)\sin^2\theta,
 \qquad x^2+y^2+z^2-a^2-2\ii az=(r-\ii a\cos\theta)^2.
 \label{eq:complexdistance}
\end{equation}
The second identity exhibits the complex distance that underlies the
Newman--Janis description.  It also explains why the coordinate roots
$r=\pm\ii a$ will occur in the separated radial equation even though the
real exterior geometry is flat.

The field strength following from \eqref{eq:potential} is nontrivial and
has the familiar electromagnetic multipoles of the Kerr--Newman black hole,
\begin{equation}
 Q_\ell+\ii P_\ell=Q(\ii a)^\ell,
 \qquad \ell=0,1,2,\ldots,
 \label{eq:emultipoles}
\end{equation}
up to the conventional choice of the sign of $a$ and of the magnetic
moments.  Thus $Q_0=Q$, $P_1=Qa$, $Q_2=-Qa^2$, and so on.  Equation
\eqref{eq:emultipoles} is the electromagnetic analogue of the multipole
relation of the Kerr black hole and is why this background is valuable in all-spin
constructions \cite{Vines:2017hyw,Arkani-Hamed:2019ymq,Chung:2019yfs,
Scheopner:2023rzp,Cangemi:2023ysz}.
The field is stationary and axisymmetric, but it is not globally source-free.

\paragraph{The disk, its two faces, and the rim.}

The oblate coordinates make the source locus geometrically transparent.
At $r=0$, equation \eqref{eq:cartesianmap} gives
\begin{equation}
 z=0,\qquad R\equiv\sqrt{x^2+y^2}=a\sin\theta,\qquad 0\leq R\leq a.
 \label{eq:diskembedding}
\end{equation}
The coordinate surface $r=0$ is therefore the equatorial disk.  It is
double covered: values $\theta$ and $\pi-\theta$ represent the two limiting
faces of the same point with $R<a$.  Approaching it from $r>0$ or $r<0$ is
not a choice of radial origin; it is a choice of face.  This point matters
for Maxwell matching, because the physical variables are the tangential and
normal field traces on the two faces, not a single scalar value of a radial
function at the coordinate value $r=0$.

The boundary $R=a$ is the ring $z=0$, $x^2+y^2=a^2$.  The angular velocity
appearing below implies a tangential speed $v=R/a$ for the material pattern:
the disk is subluminal at every $R<a$ but becomes null in the limiting sense
at the rim.  This extreme kinematics shows that the distributional source
cannot be replaced by a smooth rigid body without specifying a regulator.
It is also the geometric origin of the distinction between the ordinary
point $r=0$ of the radial CHE equation and the physical edge of the object.

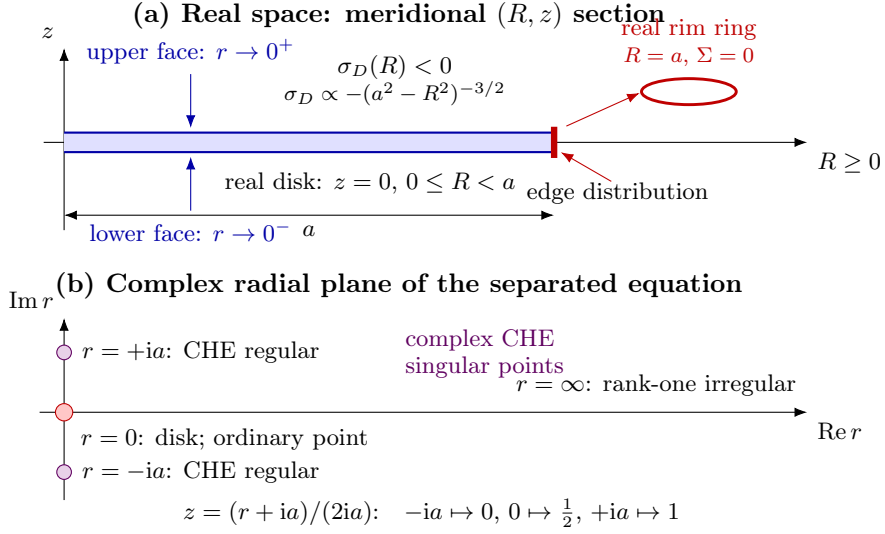
\begin{figure}[t]
\centering
\begin{tikzpicture}[x=1.08cm,y=0.86cm,>=Latex,
  every node/.style={font=\footnotesize},
  chepoint/.style={circle,draw=violet!75!black,fill=violet!18,
                   minimum size=5.5pt,inner sep=0pt},
  realpoint/.style={circle,draw=red!75!black,fill=red!22,
                    minimum size=6.5pt,inner sep=0pt}]
\begin{scope}[shift={(0,0)}]
  \node[font=\small\bfseries] at (4.1,1.95)
    {(a) Real space: meridional $(R,z)$ section};
  \draw[->] (-0.25,0) -- (9.10,0) node[below right] {$R\geq0$};
  \draw[->] (0,-1.35) -- (0,1.45) node[above left] {$z$};
  \fill[blue!12] (0,0.15) rectangle (6,-0.15);
  \draw[blue!65!black,line width=.8pt] (0,0.15) -- (6,0.15);
  \draw[blue!65!black,line width=.8pt] (0,-0.15) -- (6,-0.15);
  \draw[red!75!black,line width=2.4pt] (6,0.24) -- (6,-0.24);
  \draw[->,blue!65!black] (1.55,1.05) -- (1.55,0.22);
  \node[above,blue!65!black] at (1.55,1.05)
    {upper face: $r\to0^+$};
  \draw[->,blue!65!black] (1.55,-1.05) -- (1.55,-0.22);
  \node[below,blue!65!black] at (1.55,-1.05)
    {lower face: $r\to0^-$};
  \node[align=center] at (4.05,0.95)
    {$\sigma_D(R)<0$\\[-2pt]\scriptsize $\sigma_D\propto-(a^2-R^2)^{-3/2}$};
  \node[align=center] at (3.75,-0.63)
    {real disk: $z=0$, $0\leq R<a$};
  \node[align=left] at (6.78,-0.78)
    {edge distribution};
  \draw[->,red!75!black] (6.70,-0.60) -- (6.08,-0.15);
  \draw[->,red!75!black] (6.15,0.25) -- (7.05,0.76);
  \draw[red!75!black,line width=1.1pt] (7.65,0.78) ellipse [x radius=.58,y radius=.20];
  \node[above,align=center,red!75!black] at (7.65,1.02)
    {real rim ring\\[-2pt]\scriptsize $R=a$, $\Sigma=0$};
  \draw[<->] (0,-1.12) -- (6,-1.12) node[midway,below] {$a$};
\end{scope}
\begin{scope}[shift={(0,-4.15)}]
  \node[font=\small\bfseries] at (4.1,1.95)
    {(b) Complex radial plane of the separated equation};
  \draw[->] (-0.30,0) -- (9.10,0) node[below right] {$\operatorname{Re}r$};
  \draw[->] (0,-1.35) -- (0,1.45) node[above left] {$\operatorname{Im}r$};
  \node[realpoint] (disk) at (0,0) {};
  \node[chepoint] (minus) at (0,-0.92) {};
  \node[chepoint] (plus) at (0,0.92) {};
  \node[below right,align=left] at (disk.south east)
    {$r=0$: disk; ordinary point};
  \node[right,align=left] at (plus.east)
    {$r=+\ii a$: CHE regular};
  \node[right,align=left] at (minus.east)
    {$r=-\ii a$: CHE regular};
  \node[align=left,text=violet!75!black] at (5.15,0.90)
    {complex CHE\\[-2pt]singular points};
  \node[align=left] at (7.25,0.42)
    {$r=\infty$: rank-one irregular};
  \node[align=center] at (4.50,-1.55)
    {$z=(r+\ii a)/(2\ii a)$:\quad
     $-\ii a\mapsto0$, $0\mapsto\tfrac12$, $+\ii a\mapsto1$};
\end{scope}
\end{tikzpicture}
\caption{Physical and analytic loci associated with the root--Kerr object.
(a) shows the real meridional disk and its compensating rim
distribution. (b) shows the complex radial plane, with regular
singular points $r=\pm\ii a$ and the ordinary disk point $r=0$.
The source structure follows Lynden--Bell magic field~\cite{Lynden-Bell:2002dvr}.}
\label{fig:rootkerr-loci}
\end{figure}

\paragraph{Distributional source and its limitations.}

The source is not a smooth charged body.  In the inertial frame its disk
surface charge and azimuthal current are \cite{Lynden-Bell:2002dvr}
\begin{equation}
 \sigma_D(R)=-\frac{Qa}{2\pi(a^2-R^2)^{3/2}},\qquad
 J_\phi(R)=\frac{R}{a}\sigma_D(R),\qquad 0\leq R<a .
 \label{eq:magic-source}
\end{equation}
The divergent integrated disk contribution is compensated by an oppositely
signed distributional edge contribution at the rim, leaving total charge $Q$.  Equation
\eqref{eq:magic-source} also fixes the rigid angular velocity
\begin{equation}
 \Omega_D=\frac1a .
 \label{eq:omegaD}
\end{equation}
It follows directly that $J_\phi=\Omega_D R\sigma_D$, so the charge and
current are not independent material inputs.  They are the background
distribution required by the magic field.  What they do \emph{not} specify
is the perturbation $\delta J^\mu$ produced by an incident wave.  In
particular, the static field does not tell us whether a perturbation should
be reflected, absorbed, transmitted between faces, or converted between
helicities at the rim.

At the macroscopic level, a background charge density determines one Maxwell
solution, whereas scattering requires
a constitutive relation or an appropriate microscopic completion.  Here the
need is sharper because the disk density diverges as
$(a^2-R^2)^{-3/2}$ and the rim has an essential compensating contribution.
The local force-free or perfect-conductor condition often imposed in
stationary magnetospheres would give a different model: it is a kinematic
condition that can scatter already at order $Q^0$, while the amplitude
matching considered below calls for a response beginning at $Q^2$.  The
disk-plus-rim structure must therefore be retained when posing the photon
scattering problem, but it cannot by itself be promoted to a dynamical
boundary law.

\paragraph{A regulated-rim constraint.}

There is a sharp consequence of the stationary source which is independent
of any perturbative model.  We truncate the disk density in
\eqref{eq:magic-source} at $R=b<a$ and put
\begin{equation}
 u=a^2-b^2>0 .
 \label{eq:rim-regulator-u}
\end{equation}
This regulator should not be confused with the NS accessory parameter denoted
by $u$ below.
The retained disk then carries
\begin{align}
 Q_D(b)&=-Q\left(\frac{a}{\sqrt u}-1\right),
 \label{eq:cut-disk-charge}\\
 \mu_D(b)&=\pi\int_0^b \dd R\,R^2J_\phi(R)
 =-\frac Q2\left(\frac{a^2}{\sqrt u}+\sqrt u-2a\right).
 \label{eq:cut-disk-moment}
\end{align}
If the removed edge were replaced by one charged ring of charge $q_{\rm r}$
and current $I_{\rm r}$, matching the total charge and the magic-field
magnetic moment, $Q_{\rm tot}=Q$ and $\mu_{\rm tot}=Qa$, would require
\begin{equation}
 q_{\rm r}=\frac{Qa}{\sqrt u},\qquad
 I_{\rm r}=\frac{Q(a^2+u)}{2\pi b^2\sqrt u}.
 \label{eq:one-ring-match}
\end{equation}
For a one-component ring, $v_{\rm r}=2\pi bI_{\rm r}/q_{\rm r}$, and hence
\begin{equation}
 v_{\rm r}=\frac{a^2+u}{ab}>1.
 \label{eq:one-ring-superluminal}
\end{equation}
Thus a single subluminal corotating rim charge cannot realize even this
regulated background.  A subluminal charged rim corotating with
$\Omega_D=1/a$ supplies $I_{\rm cor}=Q/(2\pi\sqrt u)$; an independent
magnetization/current channel with
\begin{equation}
 I_{\rm mag}=\frac{Q\sqrt u}{\pi b^2},
 \qquad \mu_{\rm mag}=Q\sqrt u
 \label{eq:rim-magnetization-channel}
\end{equation}
supplies the remainder.  Any finite-speed microscopic completion must
therefore contain at least these two rim sectors (or an equivalent
counter-streaming realization).  This is a constraint from the stationary
field, not an assumed dissipative boundary condition.

The regulated integrals and the separation between charge and magnetization
channels are given in appendix~\ref{app:rimregulator}.

\section{Scalar and electromagnetic perturbation equations}

\subsection{Massless charged scalar on the root--Kerr exterior}
\label{sec:scalar}

The scalar sector isolates the distinction between a bulk coupling to the
magic field and a response of its source.  We consider a \emph{massless test
scalar} $\Phi$ of charge $q$.  On the source-free exterior it
obeys
\begin{equation}
 \bigl(\nabla_\mu-\ii q\bar A_\mu\bigr)
 \bigl(\nabla^\mu-\ii q\bar A^\mu\bigr)\Phi=0 .
 \label{eq:chargedKG}
\end{equation}
This is a probe equation: $q$ denotes the electric charge of the scalar.
There is no analogous test charge for an electromagnetic (Maxwell)
perturbation, and the $qQ$ dependence derived below must not be transferred
to that problem.
The neutral massless scalar is the consistent $q=0$ specialization.  A mass
term is not included in this work.

The magic potential in \eqref{eq:potential} is in Lorenz gauge on the
exterior, $\nabla_\mu\bar A^\mu=0$, and satisfies
\begin{equation}
 \bar A_\mu\bar A^\mu=-\frac{Q^2r^2}{\Sig\Del}.
 \label{eq:scalarAsquare}
\end{equation}
Consequently \eqref{eq:chargedKG} can be evaluated without making any
Newman--Janis prescription for a wave equation.  With the separated ansatz
\begin{equation}
 \Phi=e^{-\ii\omega t+\ii m\phi}R_{\ell m}(r)S_{\ell m}(\theta),
 \qquad m\in\mathbb Z,
 \label{eq:scalaransatz}
\end{equation}
the scalar equation separates into
\begin{align}
 \frac1{\sin\theta}\frac{\dd}{\dd\theta}
 \left(\sin\theta\frac{\dd S_{\ell m}}{\dd\theta}\right)
 +\left[a^2\omega^2\cos^2\theta-\frac{m^2}{\sin^2\theta}
 +A^{(0)}_{\ell m}(a\omega)\right]S_{\ell m}&=0,
 \label{eq:scalarangular}\\
 \frac{\dd}{\dd r}\left(\Del\frac{\dd R_{\ell m}}{\dd r}\right)
 +\left[\omega^2r^2-2qQ\omega r
 +\frac{(am+qQr)^2}{\Del}-A^{(0)}_{\ell m}(a\omega)\right]R_{\ell m}&=0 .
 \label{eq:scalarradialdirect}
\end{align}
The intermediate operator reduction and the coefficient matching to the NS
dictionary are displayed in appendix~\ref{app:separationdictionary}.
Here $A^{(0)}_{\ell m}(0)=\ell(\ell+1)$.  The term linear in $qQ$
comes from the covariant derivative, while the last numerator in
\eqref{eq:scalarradialdirect} combines the azimuthal coupling and
$-q^2\bar A^2$.  This direct derivation provides a check on the
Coulomb limit; when $a\to0$, it reduces to
\begin{equation}
 \frac{\dd}{\dd r}\left(r^2\frac{\dd R}{\dd r}\right)
 +\left[\omega^2r^2-2qQ\omega r+q^2Q^2-\ell(\ell+1)\right]R=0 .
 \label{eq:scalarcoulomblimit}
\end{equation}

\paragraph{Kerr--Newman black hole form and confluent-Heun structure.}

We define
\begin{equation}
 K_q(r)=\omega(r^2+a^2)-am-qQr,
 \qquad
 \lambda_0=A^{(0)}_{\ell m}+a^2\omega^2-2am\omega .
 \label{eq:scalarKlambda}
\end{equation}
Equation \eqref{eq:scalarradialdirect} is then equivalently
\begin{equation}
 \frac{\dd}{\dd r}\left(\Del\frac{\dd R}{\dd r}\right)
 +\left[\frac{K_q^2}{\Del}-\lambda_0\right]R=0 .
 \label{eq:scalarradial}
\end{equation}
This form makes the relation to the scalar radial operator of the
Kerr--Newman black hole transparent, while emphasizing that the present
metric is flat.  Its only
finite singularities are the complex roots $r=\pm\ii a$ of $\Del$; infinity
is rank-one irregular.  Thus it is a confluent-Heun equation.  The standard
coordinate
\begin{equation}
 z=\frac{r+\ii a}{2\ii a}
 \label{eq:scalarzmap}
\end{equation}
maps $r=-\ii a,+\ii a,0$ to $z=0,1,1/2$, respectively.  In particular, the
disk is not a radial singular point.

The local indices are
\begin{equation}
 r=-\ii a:\quad p=\pm\frac{m-\ii qQ}{2},
 \qquad
 r=+\ii a:\quad p=\pm\frac{m+\ii qQ}{2}.
 \label{eq:scalarindices}
\end{equation}
After factoring these powers, \eqref{eq:scalarradial} is in the standard
two-regular-one-irregular CHE class.  A convenient $N_f=3$
Nekrasov--Shatashvili parameterization of this equation is
\cite{Nekrasov:2009rc,Bonelli:2021uvf,
Bonelli:2022ten,Bautista:2023sdf}
\begin{align}
 a_{{\rm NS},0}^{(0)}&=\frac{m-\ii qQ}{2},&
 a_{{\rm NS},1}^{(0)}&=\frac{m+\ii qQ}{2},&
 m_3^{(0)}&=-\ii qQ,\nonumber\\
 L&=4a\omega,&
 u^{(0)}&=-\lambda_0+q^2Q^2-
 \frac{L}{2}(m+\ii qQ).
 \label{eq:scalarNSdictionary}
\end{align}
Equivalently, the two finite regular-pole masses are
$m_1=a_{{\rm NS},0}^{(0)}+a_{{\rm NS},1}^{(0)}=m$ and
$m_2=a_{{\rm NS},1}^{(0)}-a_{{\rm NS},0}^{(0)}=\ii qQ$.
The coefficient $m_3^{(0)}$ is fixed by the Coulomb power at infinity:
$R\sim r^{-1\mp\ii qQ}e^{\pm\ii\omega r}$.  
The scalar Matone branch connected to the partial wave $\ell$ is
\begin{equation}
 \alpha^{(0)}_{\ell m}(L)=
 -\nu_\ell+O(L),\qquad
 \nu_\ell\equiv
 \sqrt{\left(\ell+\frac12\right)^2-q^2Q^2}.
 \label{eq:scalarMatoneroot}
\end{equation}
The square root in \eqref{eq:scalarMatoneroot} is required by the
$q^2Q^2/r^2$ term in \eqref{eq:scalarcoulomblimit}.  In particular, it
must not be replaced by $\ell+1/2$ at nonzero $qQ$.

\paragraph{Exact exterior scalar connection coefficient.}

We denote the two local scalar CHE solutions at $z=1$ by
$\cF^{(1,0)}_\theta$, and the irregular basis at infinity by
$\cF^{(\infty,0)}_{\theta'}$, with $\theta'=+$ ($-$) the incoming
(outgoing) wave for $L>0$.
The exact scalar connection matrix takes the NS form
\cite{Bonelli:2022ten,Bautista:2023sdf}
\begin{equation}
 \mathcal C^{(0)}(\theta,\theta';\ell,m,qQ,L)
 =\left.\cM(\theta,\theta')\right|_{
 \substack{a_{{\rm NS},0}=(m-\ii qQ)/2,\;
 a_{{\rm NS},1}=(m+\ii qQ)/2\\
 m_3=-\ii qQ,\;
 u=-\lambda_0+q^2Q^2-L(m+\ii qQ)/2\\
 \alpha=\alpha^{(0)}_{\ell m}(L)}} .
 \label{eq:scalarconnection}
\end{equation}
If $c_\theta$ are the coefficients of a local scalar Cauchy solution, its
exact exterior in/out coefficients are
\begin{equation}
 b_{\rm in}^{(0)}=\sum_{\theta=\pm}c_\theta\mathcal C^{(0)}(\theta,+),
 \qquad
 b_{\rm out}^{(0)}=\sum_{\theta=\pm}c_\theta\mathcal C^{(0)}(\theta,-).
 \label{eq:scalarinout}
\end{equation}
Equations \eqref{eq:scalarconnection}--\eqref{eq:scalarinout} are the
compact NS result that will be used for scalar propagation.  A scalar
reflection coefficient is obtained only after a disk/rim condition selects
$c_\theta$.

As a nontrivial check, we take $a\to0$ while using the usual regular-origin
condition of the Coulomb problem.  We define
\begin{equation}
 \ell_{\rm eff}=-\frac12+\nu_\ell,
 \label{eq:ellEff}
\end{equation}
the standard free-$\ell$ normalization gives
\begin{equation}
 S^{\rm Coul}_{\ell}
 =e^{\ii\pi(\ell-\ell_{\rm eff})}
 \frac{\Gamma(\ell_{\rm eff}+1+\ii qQ)}
      {\Gamma(\ell_{\rm eff}+1-\ii qQ)} .
 \label{eq:scalarCoulombS}
\end{equation}
At $qQ=0$, this reduces to unity.  The formula is a check of the exterior
scalar normalization in the point-Coulomb limit; it is not a disk/rim
boundary prescription at finite $a$.

\paragraph{What the scalar connection problem determines.}

The scalar equation determines exact exterior continuation between the
ordinary disk point and an in/out basis at infinity.  For $q\ne0$, this
continuation contains the physical Coulomb-type $qQ$ phase already in the
bulk operator.  It does not, by itself, specify how a test scalar interacts
with, is transmitted through, or is reflected by the distributional disk and
rim.  Such a question requires a scalar coupling to a chosen regulated source
model.  The exterior scalar connection coefficient is therefore a rigorous
and valuable benchmark for the NS construction, but it is not automatically
an on-shell Compton amplitude of the root--Kerr object.  This qualification
will be essential when comparing it with the Maxwell problem.

\subsection{Exterior electromagnetic perturbation equation}
\label{sec:photon}

There are three distinct equations one may write down in this background, and
they should not be conflated.  First, a charged scalar on a prescribed magic
potential obeys a minimally coupled Klein--Gordon equation.  Its radial
operator contains the probe charge $q$ and the background charge $Q$ through
$q\bar A$.  Second, the field of a Maxwell perturbation on the source-free
exterior obeys the homogeneous linearized Maxwell equation.  This is the
spin-$1$ Teukolsky/CHE problem studied in this section.  Third, the physical
scattering problem includes the distributional disk/rim source and hence a
law for the induced current.  Only the third problem contains the
charge-dependent photon Compton response that can be compared with an
amplitude.

For Kerr black holes the distinction between the second and third steps is
often hidden by the availability of a canonical horizon prescription:
regularity on the future horizon selects the ingoing branch of the radial
solution.  Root--Kerr has no such locus.  The roots of $r^2+a^2$ are complex,
and the real point $r=0$ is locally ordinary for the separated equation.
The exterior ordinary differential equation (ODE) therefore determines a propagation and basis-conversion
matrix, not a reflection coefficient by itself.  This is analogous to wave
propagation in a cavity with an unspecified impedance: the bulk Green
function is well defined, but the scattering matrix is not fixed until the
boundary response is supplied.

This observation also fixes the role of the Newman--Penrose (NP) variables.  They
provide a covariant organization of the two radiative Maxwell degrees of
freedom on a principal tetrad \cite{Newman:1961qr}, as in the Teukolsky
analysis \cite{Teukolsky:1973ha,Bautista:2021wfy}.  They do not replace Maxwell reconstruction at a
material interface.  In particular, a boundary condition must act on
gauge-invariant electric and magnetic traces, not on either radiative NP
scalar in isolation.  The matching and response construction below makes this
completion explicit.

We start from Maxwell theory with an external current and perturb

\begin{equation}
 A_\mu=\bar A_\mu+a_\mu,\qquad
 f_{\mu\nu}=2\partial_{[\mu}a_{\nu]} .
\end{equation}
Because Maxwell theory is quadratic, its exact linearization on the fixed
root--Kerr geometry takes the form
\begin{equation}
 \nabla_\nu f^{\mu\nu}=\delta J^\mu,\qquad
 \nabla_{[\mu}f_{\nu\rho]}=0 .
 \label{eq:linmaxwell}
\end{equation}
Away from the source, $\delta J^\mu=0$.  Neither $\bar A_\mu$ nor $Q$
appears in \eqref{eq:linmaxwell}.  This is the central distinction from a
charged scalar probe: the latter has a minimal $q\bar A$ coupling, whereas a
Maxwell perturbation of an Abelian background does not.  The $Q^2$ dependence of a physical photon Compton
amplitude must consequently be supplied by the disk/rim response.

\paragraph{Newman--Penrose derivation and separation.}

A principal tetrad is
\cite{Kinnersley:1969zza,Teukolsky:1973ha}
\begin{align}
 \ell^\mu\partial_\mu&=\partial_t+\partial_r+
 \frac{a}{\Del}\partial_\phi,\\
 n^\mu\partial_\mu&=\frac{r^2+a^2}{2\Sig}\partial_t
 -\frac{\Del}{2\Sig}\partial_r+\frac{a}{2\Sig}\partial_\phi,\\
 m^\mu\partial_\mu&=\frac{1}{\sqrt2\,(r+\ii a\cos\theta)}
 \left(\ii a\sin\theta\,\partial_t+\partial_\theta+
 \frac{\ii}{\sin\theta}\partial_\phi\right).
 \label{eq:tetrad}
\end{align}
For the two radiative Maxwell scalars, we define
\begin{equation}
 \psi_{+1}=\phi_0,\qquad
 \psi_{-1}=\rho^{-2}\phi_2,\qquad
 \rho=-\frac1{r-\ii a\cos\theta} .
 \label{eq:teukvars}
\end{equation}
Eliminating $\phi_1$ from the NP equations gives the source-free spin-$s$
Teukolsky equation.  Equivalently, it is the controlled $M\to0$ limit of
the Teukolsky equation of the Kerr black hole \cite{Teukolsky:1973ha}.  With
\begin{equation}
 \psi_s=e^{-\ii\omega t+\ii m\phi}R_s(r)\,{}_sS_{\ell m}(\theta),
 \qquad s=\pm1,
 \label{eq:separation}
\end{equation}
we define
\begin{equation}
 K=\omega(r^2+a^2)-am,\qquad
 \lambda_s={}_sA_{\ell m}(a\omega)+a^2\omega^2-2am\omega .
 \label{eq:defKlambda}
\end{equation}
The radial equation is
\begin{equation}
 \Del^{-s}\frac{\dd}{\dd r}\left(\Del^{s+1}\frac{\dd R_s}{\dd r}\right)
 +\left[\frac{K^2-2\ii srK}{\Del}+4\ii s\omega r-\lambda_s\right]R_s=0 .
 \label{eq:radial}
\end{equation}
Appendix~\ref{app:npdecoupling} gives the decoupling identity that leads from
the first-order NP Maxwell equations to \eqref{eq:radial} and
\eqref{eq:angular}.
The angular equation is the usual spin-weighted spheroidal equation,
\begin{align}
 \frac1{\sin\theta}\frac{\dd}{\dd\theta}
 \left(\sin\theta\frac{\dd{}_sS}{\dd\theta}\right)
 +\bigg[&a^2\omega^2\cos^2\theta-2a\omega s\cos\theta+s+{}_sA_{\ell m}
 \nonumber\\[-4pt]
 &-\frac{(m+s\cos\theta)^2}{\sin^2\theta}\bigg]{}_sS=0 .
 \label{eq:angular}
\end{align}
At $a\omega=0$, ${}_sA_{\ell m}=\ell(\ell+1)-s(s+1)$.  The quantity that
appears in the angular operator \eqref{eq:angular} is
$s+{}_sA_{\ell m}$; for either radiative spin weight $s=\pm1$, it is therefore
$\ell(\ell+1)-1$.  This fixes the spherical-limit normalization of the
electromagnetic radial equation.

The construction above clarifies the role of Newman--Janis.  It correctly
generates the background, a principal tetrad, and the locations of the
complex radial singularities.  It is not a rule for transforming a separated
perturbation equation term by term.  Such a procedure misses derivatives,
spin coefficients, tetrad phases and the field redefinition in
\eqref{eq:teukvars}; in the present setting it would also incorrectly retain
a scalar $qQ$ coupling in the electromagnetic radial equation.

\paragraph{NP decoupling checks.}
\label{sec:npdetails}

The nonzero spin coefficients of \eqref{eq:tetrad} needed to decouple the
radiative Newman--Penrose Maxwell scalars $\phi_0$ and $\phi_2$ are
\cite{Newman:1961qr,Kinnersley:1969zza,Teukolsky:1973ha}
\begin{align}
 \rho&=-\frac1{r-\ii a\cos\theta},
&\mu&=-\frac{\Del}{2\Sig(r-\ii a\cos\theta)},
&\tau&=-\frac{\ii a\sin\theta}{\sqrt2\Sig},\nonumber\\
 \pi&=\frac{\ii a\sin\theta}{\sqrt2(r-\ii a\cos\theta)^2},
&\beta&=\frac{\cot\theta}{2\sqrt2(r+\ii a\cos\theta)},
&\alpha&=\pi-\bar\beta,\nonumber\\
 \gamma&=\mu+\frac{r}{2\Sig},&\varepsilon&=0 .
 \label{eq:spincoefficients}
\end{align}
The principal congruences obey
\begin{equation}
 \kappa=\sigma=\lambda=\nu=0 .
 \label{eq:geodesicshearfree}
\end{equation}
The two Maxwell equations that eliminate $\phi_1$ in the $s=+1$ sector are
\begin{align}
 (D-2\rho)\phi_1-(\bar\delta+\pi-2\alpha)\phi_0&=0,\nonumber\\
 (\delta-2\tau)\phi_1-(\Delta+\mu-2\gamma)\phi_0&=0 .
 \label{eq:nppair}
\end{align}
We apply $(\delta-2\tau)$ to the first equation and $(D-2\rho)$ to the
second and use the NP commutators to eliminate $\phi_1$.  The corresponding
pair for $\rho^{-2}\phi_2$ yields $s=-1$.  The background Maxwell scalar
$\bar\Phi_1$ never enters this operation: the geometry and the tetrad are
held fixed, while \eqref{eq:linmaxwell} is the homogeneous Maxwell equation
for $f$.

Three independent checks constrain the result.  First, we write
$R_+\equiv R_{s=+1}$ and $R_-\equiv R_{s=-1}$.  In the nonrotating
(spherical) limit $a\to0$, the
separated radial equations become
\begin{align}
 r^{-2}\frac{\dd}{\dd r}\left(r^4R_{+}'\right)
 +\left[\omega^2r^2-{}_1A_{\ell m}(0)+2\ii\omega r\right]R_+&=0,
 \nonumber\\
 r^2R_-''+\left[\omega^2r^2-{}_{-1}A_{\ell m}(0)-2\ii\omega r\right]R_-&=0.
 \label{eq:sphericallimit}
\end{align}
These are equations for the radiative NP Maxwell scalars, not for a scalar
radial function; they should not be compared to the scalar Coulomb equation
after a naive $R\mapsto R/r$ substitution.  Second, the local radial exponents are
\begin{align}
 r=-\ii a:&\qquad p=\frac m2,\quad-\frac m2-s,\nonumber\\
 r=+\ii a:&\qquad p=\frac m2-s,\quad-\frac m2 .
 \label{eq:localexponents}
\end{align}
After removing the common shift $-s/2$, these reproduce the monodromies in
\eqref{eq:nsmonodromy}.  Finally, taking the controlled $M\to0$ limit of
the Teukolsky equation for the Kerr black hole reproduces both
\eqref{eq:radial} and \eqref{eq:spincoefficients}, providing an independent
check of the direct Newman--Penrose derivation.

\subsection{Exterior CHE propagation and disk/rim matching}
\label{sec:ns}

\paragraph{Analytic role of the confluent-Heun problem.}

The appearance of a confluent Heun equation is not, by itself, a boundary
condition or an amplitude calculation.  Its value is that it organizes the
global analytic continuation of the separated exterior equation.  At an
ordinary point $z_*$, a Cauchy basis is the fundamental pair $u_1,u_2$ fixed
by $u_1(z_*)=1$, $u_1'(z_*)=0$ and $u_2(z_*)=0$, $u_2'(z_*)=1$.  This basis,
a local basis near either regular singular point, and the in/out basis at
infinity are all related by finite connection matrices.  For the Kerr black
hole this structure is used together with a horizon boundary
condition to obtain partial-wave data.  In root--Kerr the same structure
survives, but the final row vector selecting a physical solution must instead
come from the disk/rim response.

The NS description provides a practical nonperturbative parameterization of
these connection coefficients.  In the three-flavour ($N_f=3$) gauge-theory
dictionary, the accessory parameter---the coefficient not fixed by the local
singularity exponents---is determined by the Matone relation, while derivatives
of the NS free energy control the connection formula.  This representation
is particularly efficient in the low-frequency regime because one may expand
the free energy before evaluating the physical integer angular-momentum
limit.  The two-dimensional conformal-field-theory (CFT$_2$) and
instanton-counting construction of these formulae and its
application to the Kerr black hole were developed in
\cite{Bonelli:2021uvf,Bonelli:2022ten,
Bautista:2023sdf}.  Our task is to retain the exact exterior machinery while
changing the interpretation of the interior data.

Two geometrical facts keep the analytic and physical problems distinct.  The
two regular CHE points are at complex values of the oblate radial coordinate,
whereas the disk lies on the real continuation contour.  Moreover, the disk
point is ordinary for the ODE; no local monodromy can encode the material
response.  The connection matrix is thus necessary for carrying a boundary
law to infinity, but it cannot determine that law.  This is the reason for
developing the trace-space factorization below.

\paragraph{Electromagnetic dictionary.}

The roots $r=\pm\ii a$ of $\Del$ are regular singular points of
\eqref{eq:radial}, whereas infinity is rank-one irregular.  Thus it is a
confluent Heun equation.  We set
\begin{equation}
 z=\frac{r+\ii a}{2\ii a} .
 \label{eq:zmap}
\end{equation}
Then $r=-\ii a,+\ii a,\infty$ map to $z=0,1,\infty$, while the physical
disk is the ordinary point
\begin{equation}
 z_D=\frac12 .
 \label{eq:diskpoint}
\end{equation}
The local exponents of the electromagnetic radial equation \eqref{eq:radial} give the two NS monodromy
parameters
\begin{equation}
 a_{{\rm NS},0}=\frac{m+s}{2},\qquad
 a_{{\rm NS},1}=\frac{m-s}{2} .
 \label{eq:nsmonodromy}
\end{equation}
In the conventions of the NS functions used for scattering by Kerr black holes, the full
root--Kerr dictionary is
\begin{equation}
 L=4a\omega,\qquad
 (m_1,m_2,m_3)=(m,-s,-s),\qquad
 u=-\lambda_s-s(s+1)-\frac{L}{2}(m+s).
 \label{eq:dictionary}
\end{equation}
The zero-frequency branch appropriate to the partial wave $\ell$ is
\begin{equation}
 \alpha=-\ell-\frac12+O(L),\qquad
 u=\frac14-\alpha^2+L\partial_L\cF_{\rm NS} .
 \label{eq:matone}
\end{equation}
Here $\cF_{\rm NS}(m_1,m_2,m_3,\alpha,L)$ is the NS free energy.
The residue comparison that yields the scalar and electromagnetic dictionaries is given
in appendix~\ref{app:separationdictionary}.

For example, for $(s,m)=(1,1)$, the first two radial separation constants are
\begin{align}
 \lambda_{111}&=-\frac{3L}{4}+\frac{9L^2}{320}-\frac{3L^3}{2560}+O(L^4),
 \label{eq:lambda111}\\
 \lambda_{211}&=4-\frac{7L}{12}+\frac{479L^2}{12096}
 +\frac{611L^3}{870912}+O(L^4).
 \label{eq:lambda211}
\end{align}

\paragraph{Connection matrix and the physical boundary problem.}

We denote a local CHE basis at $z=1$ by $\cF^{(1)}_\theta$ and an irregular
basis at infinity by $\cF^{(\infty)}_{\theta'}$.  The exact NS
connection formula has the form \cite{Bautista:2023sdf}
\begin{equation}
 \cF^{(1)}_\theta(1-z)=\sum_{\theta'=\pm}
 \cM(\theta,\theta')\cF^{(\infty)}_{\theta'}(z^{-1}),
 \label{eq:connection}
\end{equation}
where
\begin{align}
 \cM(\theta,\theta')={}&\sum_{\sigma=\pm}L^{\sigma\alpha}
 \frac{\Gamma(1-2\sigma\alpha)\Gamma(-2\sigma\alpha)}
 {\Gamma(\frac12+\theta a_{{\rm NS},1}-\sigma\alpha+a_{{\rm NS},0})}
 \nonumber\\
 &\times\frac{\Gamma(1+2\theta a_{{\rm NS},1})}
 {\Gamma(\frac12+\theta a_{{\rm NS},1}-\sigma\alpha-a_{{\rm NS},0})
  \Gamma(\frac12-\sigma\alpha-\theta'm_3)}
 \nonumber\\
 &\times\exp\left[\ii\pi\frac{1-\theta'}2
 \left(\frac12-m_3-\sigma\alpha\right)-\frac\sigma2\partial_\alpha\cF_{\rm NS}\right].
 \label{eq:connectioncoefficient}
\end{align}
At physical integer $\ell$, the two local Frobenius exponents differ by an
integer.  The second local solution can then acquire a logarithmic term, while
individual gamma functions in the connection coefficients develop compensating
singular factors.  We define the result by analytic continuation from
non-integer $\ell$.  We introduce a temporary displacement $x$ of a gamma-function
argument that approaches a non-positive integer, so that the argument is
$-n+x$.  The regulator $x$ has no physical meaning and is taken to zero only
after the local regular/logarithmic basis has been assembled.  We then resolve
the gamma poles before expanding in $L$, using
\begin{equation}
 \Gamma(-n+x)=\frac{\Gamma(1+x)}{x\prod_{j=1}^n(x-j)},\qquad
 L^{\alpha(L)}=L^{\alpha(0)}
 \exp\left[(\alpha(L)-\alpha(0))\log L\right].
 \label{eq:regularization}
\end{equation}
The first identity displays the pole explicitly; the second keeps the
$L$-dependence of the Matone solution $\alpha(L)$ and hence generates the
required powers of $\log L$.  This gives a finite canonical connection matrix
$C^{(\ell)}$ from the local regular/logarithmic basis to the asymptotic
in/out basis.  Its rows are labelled by $\{{\rm reg},{\rm log}\}$ and its
columns by $\{{\rm in},{\rm out}\}$.  In particular, the regular row
\begin{equation}
 C^{(\ell)}_{\rm reg}\equiv
 \bigl(C^{(\ell)}_{{\rm reg},{\rm in}},
       C^{(\ell)}_{{\rm reg},{\rm out}}\bigr)
 \sim L^{-\ell-1/2}
 \label{eq:connectiontower}
\end{equation}
has this leading behavior componentwise, up to finite coefficients.  It is a
normalization of the local basis, not a physical scattering amplitude; the
second row contains the expected logarithms.  We use the NS free energy through
$L^6$ to construct $C^{(\ell)}$ for $(s,m)=(1,1)$ and every
$\ell=1,\ldots,7$.

Equation \eqref{eq:connection} is exact exterior propagation data.  It is
not yet a physical scattering amplitude.  To make the propagation map
explicit, we define the disk evaluation matrix
\begin{equation}
 \cE_D\equiv\left.
 \begin{pmatrix}
 \cF^{(1)}_+(1-z)&\cF^{(1)}_-(1-z)\\
 \partial_z\cF^{(1)}_+(1-z)&\partial_z\cF^{(1)}_-(1-z)
 \end{pmatrix}\right|_{z=z_D} .
 \label{eq:diskevaluation}
\end{equation}
Thus $b^{(1)}=\cE_D^{-1}Y_D$ are the coefficients of the disk Cauchy vector
$Y_D=(R_s(z_D),\partial_zR_s(z_D))^T$ for a fixed $(s,\ell,m)$ channel in
the local $z=1$ basis.  The coefficients
in the asymptotic in/out basis are consequently
\begin{equation}
 c_\infty=\cM^T b^{(1)}=\cM^T\cE_D^{-1}Y_D .
 \label{eq:cauchy}
\end{equation}
The transpose follows from the index ordering in \eqref{eq:connection}:
the local-basis label is the first index of $\cM$.  The physical source
response fixes $Y_D$ and mixes helicities and angular
multipoles.  Neither $z=0$ nor $z=1$ lies on the physical real radial line or corresponds to a horizon. Consequently, a Kerr-style ingoing horizon condition cannot be imposed at either finite CHE singular point.

\paragraph{Disk/rim matching at an ordinary point.}
\label{sec:diskmatching}

The oblate coordinates are related to cylindrical coordinates by
\begin{equation}
 R=\sqrt{r^2+a^2}\sin\theta,\qquad Z=r\cos\theta .
 \label{eq:oblate}
\end{equation}
For real $r$, \eqref{eq:zmap} is the vertical line
$\operatorname{Re}z=1/2$ in the CHE plane.  The upper and lower limits
$r\to0^\pm$ represent the two faces of the same material disk, with
$R=a\sin\theta$.  Analytic continuation through $r=0$ is consequently a
coordinate continuation, not a physical boundary condition after the
distributional source is inserted.  The rim at $R=a$ must be included as a
separate boundary component.

Lynden-Bell's finite-speed rotating disks provide a regulator of the
strictly magic, null-rim limit.  At finite edge velocity,
\begin{equation}
 v_{\rm edge}=\Omega a<1,\qquad
 u^\mu=\gamma(R)\bigl((\partial_t)^\mu+\Omega(\partial_\phi)^\mu\bigr),
 \qquad \gamma(R)=\bigl(1-\Omega^2R^2\bigr)^{-1/2} .
 \label{eq:finitespeed}
\end{equation}
A regulated response problem is required before taking the
$v_{\rm edge}\to1$ limit.  The stationary magic solution itself supplies
only the one-point current \eqref{eq:magic-source}.

\paragraph{Maxwell jumps and response scaling.}

We take $n^\mu$ to point from the lower to the upper disk face.  For any
quantity with limits on the two faces, we write
$[X]\equiv X|_{r=0^+}-X|_{r=0^-}$.  The exact Maxwell interface conditions
for a surface current $j_D^\mu$ are
\begin{equation}
 n_\mu[F^{\mu\nu}]=4\pi j_D^\nu,\qquad
 n_\mu[{}^\star F^{\mu\nu}]=0,\qquad
 D_a j_D^a=0 .
 \label{eq:maxwelljumps}
\end{equation}
Here $D_a$ is the covariant derivative intrinsic to the disk worldvolume,
namely the $2+1$ dimensional surface swept out by the rotating disk.
Linearization about the magic source gives
\begin{equation}
 n_\mu[f^{\mu\nu}]=4\pi\delta j_D^\nu,\qquad
 n_\mu[{}^\star f^{\mu\nu}]=0 .
 \label{eq:linearjumps}
\end{equation}
Equations \eqref{eq:linearjumps} are not closed until a constitutive or
mechanical equation determines $\delta j_D^\nu$ and the corresponding rim
current.

Here and below, bars denote the stationary magic background:
$F_{\mu\nu}=\bar F_{\mu\nu}+f_{\mu\nu}$ and
$j_D^a=\bar j_D^a+\delta j_D^a$.
This point also explains the charge counting.  A fixed perfect conductor has
the rest-frame condition $P^\mu{}_{\nu}F^{\nu\rho}u_\rho=0$ and scatters
already at $Q^0$; it is therefore not the desired root--Kerr Compton model.
For a dynamical charged disk, we write schematically
\begin{equation}
 j_D^a=\sigma u^a+k^a,\qquad \bar j_D^a=O(Q),\qquad \bar F=O(Q).
 \label{eq:currentdecomp}
\end{equation}
If $X$ denotes disk and rim mechanical variables, their linearized force
equation has the schematic form
\begin{equation}
 \mathcal L_{D+R}\,\delta X
 =\bar j_D^a f_{\,a}+\delta\mathfrak f_{\rm rim} .
 \label{eq:forcebalance}
\end{equation}
Here $\delta\mathfrak f_{\rm rim}$ denotes the force transmitted by the
perturbed rim sectors.  This term makes explicit that the disk and rim cannot
be varied independently.
At leading order, we write the electromagnetic perturbation as
$f=f_{\rm inc}+f_{\rm scattered}$, where $f_{\rm inc}$ is a prescribed
incident source-free wave and $f_{\rm scattered}$ is the field radiated by
the induced disk/rim current.  For a finite, $Q$-independent mechanical
operator,
\begin{equation}
 \delta X=O(Qf_{\rm inc}),\qquad
 \delta j_{D+R}=O(Q^2f_{\rm inc}),\qquad
 f_{\rm scattered}=O(Q^2f_{\rm inc}) .
 \label{eq:qcounting}
\end{equation}
Thus the $Q^2$ dependence absent from the homogeneous radial equation is
precisely supplied by the dynamical source response.

\paragraph{Trace-space boundary law.}

For each frequency and azimuthal number, reconstruct the Maxwell field from
both radiative Teukolsky variables on the two disk faces.  The relevant traces
are the gauge-invariant restrictions of the electromagnetic field and its
normal flux to the disk/rim; the word ``trace'' is used in this boundary-value
sense.  A linear boundary law, if a dynamical source model is specified, has
the form
\begin{equation}
 n_\mu[f^{\mu\nu}]=4\pi Q^2
 \Pi_{D+R}^{\nu;\rho\sigma}(\omega,m)f_{\rho\sigma},
 \qquad n_\mu[{}^\star f^{\mu\nu}]=0 .
 \label{eq:kernelbc}
\end{equation}
The unknown retarded kernel $\Pi_{D+R}$ must respect current conservation on
the disk worldvolume and is generically nonlocal in $R$.  Since the corotating velocity
varies over the disk, its partial-wave projection preserves $(\omega,m)$ but
need not be diagonal in $\ell$ and can mix helicities.  Consequently, a
scalar radial condition such as $R_s(0)=0$ does not by itself determine the
physical photon amplitude.

\section{On-shell amplitude data and the disk/rim boundary problem}
\label{sec:response}

The exterior Maxwell equation and its CHE/NS connection data do not select a
scattering state: they propagate a specified disk/rim trace to infinity.  At
fixed $(\omega,m)$ the physical linearized junction condition has the form
\begin{equation}
 [f_{ab}]=0,\qquad
 n_\mu[f^{\mu\nu}]e_\nu{}^a=4\pi\delta j^a,
 \qquad D_a\delta j^a=0 .
 \label{eq:linearized-junction}
\end{equation}
It becomes a scattering condition only after a retarded relation between the
induced current and the gauge-invariant disk/rim field traces is supplied.
The stationary source in section~\ref{sec:background} does not supply this
relation.

\paragraph{On-shell Compton-amplitude data.}

The all-spin root--Kerr Compton amplitude provides a precise
on-shell target for a prospective source model.  It supplies closed all-spin
expressions in both helicity sectors \cite{Cangemi:2023ysz,Cangemi:2023bpe}.
For the opposite-helicity sector, we make the pole/local separation explicit.
It identifies the analytic part of the Compton amplitude that must be matched
by the effective disk/rim response in the chosen factorization scheme.  In
crossed kinematics, $p$ denotes the massive-source momentum and $k$ and
$k'$ the two photon momenta.  The vector
$\chi^\mu=\langle3|\sigma^\mu|4]$ is the standard massless
spinor-helicity bilinear for the two photon legs.  We take
\begin{equation}
 q_\perp=k+k',\qquad P:=p\!\cdot q_\perp=2M\omega,\qquad
 \chi^\mu:=\langle3|\sigma^\mu|4],
 \label{eq:crossedvariables}
\end{equation}
We then introduce the abbreviations
\begin{equation}
 C:=p\!\cdot\chi,\qquad A:=a\!\cdot\chi,\qquad
 x:=a\!\cdot q_\perp,\qquad y:=a\!\cdot(k'-k),\qquad
 z:=2a\omega,\qquad w:=\frac{AP}{C}.
 \label{eq:localvariables}
\end{equation}
Thus, $A$, $C$, and $P$ are abbreviations introduced here; they are not
additional response parameters.  In this notation the selected all-spin
amplitude is the sum of a canonical massive-source pole representative and
an analytic remainder,
\begin{align}
 \mathcal M^{-+}_{\rm full}
 &=\frac{2Q^2C^2}{P^2}
 \left[e^x\cosh z-w e^x\operatorname{sinhc}z
 +\frac12\left(w^2-z^2\right)E(x,y,z)\right],
 \label{eq:fullflipsplit}\\
 \mathcal M^{-+}_{\rm pole}
 &=\frac{2Q^2C^2}{P^2}
 \left[e^x\cosh z-w e^x\operatorname{sinhc}z\right],
 \qquad \operatorname{sinhc}z:=\frac{\sinh z}{z}.
 \label{eq:flipcanonicalpole}
\end{align}
The first line is the crossed form of the amplitude input, while the second
line specifies the subtraction scheme.  Substituting $w=AP/C$ gives
\begin{equation}
 \mathcal M_{\rm loc}^{-+}:=\mathcal M_{\rm full}^{-+}
 -\mathcal M_{\rm pole}^{-+}
 =Q^2\left(A^2-\frac{C^2z^2}{P^2}\right)E(x,y,z).
 \label{eq:localamplitude}
\end{equation}
The apparently nonlocal factor $w$ has therefore cancelled.  The remaining
function $E$ is entire in $(x,y,z)$ and contains the spin-induced local
completion.  This pole/local separation is convention dependent only in the
usual analytic contact-term sense; equations \eqref{eq:flipcanonicalpole}
and \eqref{eq:localamplitude} define the convention adopted here.

To pass from this on-shell expression to a partial-wave kernel, we use the
unit-flux projection
\begin{equation}
 [\cK^{m}]_{\ell'\lambda',\ell\lambda}
 =\frac{\omega}{16\pi^2 M Q^2}
 \int\dd\Omega'\dd\Omega\,
 {}_{-\lambda'}Y_{\ell'm}^{*}(\Omega')\,
 \mathcal M^{\rm crossed}_{\lambda'\lambda}(\Omega',\Omega)\,
 {}_{-\lambda}Y_{\ell m}(\Omega).
 \label{eq:localprojection}
\end{equation}
Thus $\cK^m$ is, by definition, the unit-flux partial-wave representation
of the Compton amplitude: its row $(\ell',\lambda')$ labels an outgoing
photon mode and its column $(\ell,\lambda)$ labels an incoming mode at fixed
$m$.  It is an amplitude kernel, not yet the physical scattering matrix.
For an arbitrary partial-wave kernel $\mathsf A$, we define the linear
operators $\mathscr X$ and $\mathscr Y$ to represent multiplication by the
kinematic variables $x$ and $y$ inside the angular projection:
\begin{align}
 \mathscr X\mathsf A&=(a\omega X_{-1}^{m})\mathsf A+
 \mathsf A(a\omega X_{+1}^{m}),\nonumber\\
 \mathscr Y\mathsf A&=(a\omega X_{-1}^{m})\mathsf A-
 \mathsf A(a\omega X_{+1}^{m}),
 \label{eq:localXY}
\end{align}
where the left and right actions correspond, respectively, to the outgoing
and incoming angular variables.  The matrix $X_s^m$ is the tridiagonal
representation of multiplication by $\cos\theta$ on spin-weighted
harmonics, given explicitly in appendix~\ref{app:angular}.
The direct projection of the order-$a^2$
part of \eqref{eq:localamplitude} has support only on the $\ell=1$
helicity-flip block.  Appendix~\ref{app:quadrupoleseed} derives this
projection, including the Wigner normalization and the two angular integrals.
For $m=1$ it gives
\begin{equation}
 [\cU_{\rm loc}^{m=1}]_{1-,1+}
 =-\frac{2a^2\omega^3}{3\pi M},
 \qquad [\cU_{\rm loc}^{m=1}]_{\ell'\lambda',\ell\lambda}=0
 \quad\hbox{otherwise}.
 \label{eq:allspinlift}
\end{equation}
We write $E=\sum_{n\geq0}E_n$, with $E_n$ homogeneous of degree $n$.  The
full local lift and its expansion through order $a^8$ are consequently
\begin{equation}
 \cK_{\rm loc}^{-+}=E(\mathscr X,\mathscr Y,2a\omega)\cU_{\rm loc},
 \qquad
 [\cK_{\rm loc}^{-+}]_{\leq a^8}=
 \sum_{N=2}^{8}E_{N-2}(\mathscr X,\mathscr Y,2a\omega)\cU_{\rm loc}
 +O(a^9).
 \label{eq:lift}
\end{equation}
The scale $M$ is the massive-state normalization used in the amplitude, not
a mass parameter of the flat root--Kerr geometry.  Equation
\eqref{eq:allspinlift} follows from the explicit unit-flux projection;
equation \eqref{eq:lift} is its finite-order all-spin continuation.  Neither
is obtained by imposing a boundary condition on the Maxwell equation.

For reference, in the $m=1$ sector the first nonzero transitions from an
incoming $|\ell=1,+\rangle$ state obey the kinematical finite-band pattern
\begin{equation}
 [\cK_{\rm loc}^{m=1}]_{\ell'-,1+}=a^{\ell'+1}c_{\ell'}(\omega)
 +O(a^{\ell'+2}),\qquad \ell'=1,\ldots,7,
 \label{eq:ladderform}
\end{equation}
where
\begin{align}
 c_1&=-\frac{2\omega^3}{3\pi M},&
 c_2&=-\frac{2\omega^4}{3\sqrt{15}\pi M},&
 c_3&=-\frac{8\omega^5}{45\sqrt{21}\pi M},\nonumber\\
 c_4&=-\frac{\omega^6}{105\sqrt3\pi M},&
 c_5&=-\frac{2\omega^7}{525\sqrt{33}\pi M},&
 c_6&=-\frac{4\omega^8}{10395\sqrt{39}\pi M},\nonumber\\
 c_7&=-\frac{32\omega^9}{2837835\sqrt5\pi M}.
 \label{eq:ladder}
\end{align}
These values follow by applying the homogeneous polynomials $E_0,\ldots,E_6$
to the seed \eqref{eq:allspinlift}.  Appendix~\ref{app:allspinlift}
derives the first two entries explicitly from the tridiagonal Wigner action;
the remaining entries follow from the same finite matrix operation.  They are
matching targets, not root--Kerr predictions from the exterior CHE equation
alone.

\paragraph{What amplitude matching fixes.}

The disk/rim traces contain more data than is carried by two radiative photon
helicities.  At fixed $m$ and in a finite partial-wave regulator, we write
\begin{equation}
 \mathsf T_A:\mathscr H_m\longrightarrow\mathscr V_{\rm tr},
 \qquad A={\rm in,out},
 \label{eq:fulltracemaps}
\end{equation}
Here $\mathscr H_m$ is the regulated space of unit-flux radiative photon
partial waves at fixed $m$, while $\mathscr V_{\rm tr}$ is the larger space
of gauge-invariant electromagnetic traces on the two disk faces and at the
rim.  Thus $\mathsf T_{\rm in}$ maps an incoming photon channel to its
disk/rim traces, and $\mathsf T_{\rm out}$ does the same for an outgoing
channel.  Both maps are determined by the CHE/NS connection problem and the
Maxwell-field reconstruction.  Their images are the radiation-visible
subspaces of $\mathscr V_{\rm tr}$.

We use $\mathfrak R$ to denote the effective linear trace-to-trace response
of the disk/rim: it incorporates the induced current specified by a
microscopic version of \eqref{eq:kernelbc}, together with the associated
Maxwell jump.  The Compton amplitude constrains this response only between
the radiation-visible subspaces.  Let $\mathsf T_{\rm out}^{+}$ be a chosen
left inverse of $\mathsf T_{\rm out}$ on its image, so that it converts a
radiation-visible outgoing trace back into its outgoing photon channel.  A
response operator matches the prescribed Compton channel kernel
$\cK_{\rm C}$ precisely when
\begin{equation}
\mathsf T_{\rm out}^{+}\,\mathfrak R\,\mathsf T_{\rm in}
=\ii Q^2\cK_{\rm C}.
\label{eq:minimalreaction}
\end{equation}
Here $\cK_{\rm C}$ is the unit-flux partial-wave kernel obtained from the
selected all-spin Compton amplitude.  An explicit regulated construction of
the trace maps and their channel projection is given in
appendix~\ref{app:traceprojection}.  Equation \eqref{eq:minimalreaction}
therefore states that mapping an incident photon to the disk/rim, applying
the source response, and projecting the result back onto outgoing radiative
channels reproduces the prescribed amplitude.

This condition does \emph{not} determine the action of $\mathfrak R$ on
trace directions which no radiative photon probes.  We call a representative
that has no additional action on those unprobed directions a \emph{minimal
on-shell completion}.  It is a matching convention, not a microscopic
disk/rim law and not a prediction for dissipation, a rim spectrum, or a
static disk/rim susceptibility.  When isolating the local part of the
amplitude input, $\cK_{\rm eff}$ denotes the partial-wave kernel of the full
selected effective amplitude and $\cK_{\rm pole}$ the kernel of its pole
representative in \eqref{eq:flipcanonicalpole}; we define
\begin{equation}
\cK_{\rm loc}\equiv\cK_{\rm eff}-\cK_{\rm pole}
\label{eq:localchannelkernel}
\end{equation}
as the pole-subtracted local channel kernel.

\paragraph{From a boundary law to phase shifts.}

The same statement can be written in the familiar language of a general
inner boundary condition.  We first consider a finite channel truncation and a
choice of two disk-side local wave bases, denoted by $u_D^{\rm in}$ and
$u_D^{\rm out}$.  These labels become physical only after a normal-flux
metric and an orientation at the two-faced disk have been specified.  We
denote their exact CHE/NS continuation to infinity by
\begin{equation}
 u_D^A=\mathsf C_{{\rm in},A}\,u_\infty^{\rm in}
       +\mathsf C_{{\rm out},A}\,u_\infty^{\rm out},
 \qquad A={\rm in,out} .
 \label{eq:diskconnectionblocks}
\end{equation}
Here the matrices $\mathsf C$ are the Cauchy-to-in/out blocks obtained from
the connection data in \eqref{eq:cauchy}, including the reconstruction from
the separated radial variables to physical traces.  A general linear disk/rim
condition can then be parameterized as
\begin{equation}
 a_D^{\rm out}=\mathsf R_D(\omega,m)\,a_D^{\rm in} .
 \label{eq:general-disk-reflection}
\end{equation}
It gives the response-dressed scattering matrix
\begin{equation}
 \mathsf S_{\mathsf R_D}=
 \bigl(\mathsf C_{{\rm out},{\rm in}}
       +\mathsf C_{{\rm out},{\rm out}}\mathsf R_D\bigr)
 \bigl(\mathsf C_{{\rm in},{\rm in}}
       +\mathsf C_{{\rm in},{\rm out}}\mathsf R_D\bigr)^{-1} .
 \label{eq:response-dressed-S}
\end{equation}
Equation \eqref{eq:response-dressed-S} separates exact exterior propagation,
encoded in the connection blocks $\mathsf C$, from the unknown source
dynamics encoded in $\mathsf R_D$.  A scalar reflection coefficient is
recovered only in a diagonal, separable model.  In general $\mathsf R_D$ can
mix the two disk faces, photon helicities, and angular multipoles; a causal
passive model further constrains it through the disk flux form.

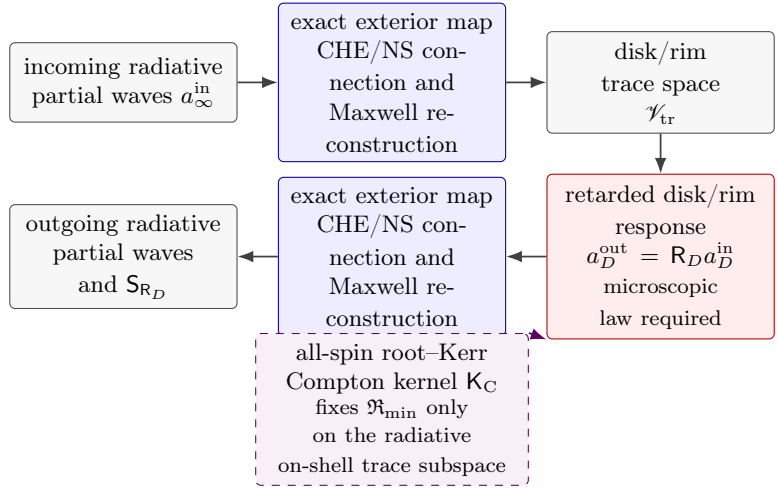
\begin{figure}[t]
\centering
\begin{tikzpicture}[x=0.96cm,y=0.94cm,>=Latex,
  every node/.style={font=\footnotesize},
  flowbox/.style={draw=black!70,rounded corners=2pt,fill=black!3,
                  align=center,text width=2.75cm,minimum height=1.05cm},
  propbox/.style={draw=blue!65!black,rounded corners=2pt,fill=blue!7,
                  align=center,text width=2.75cm,minimum height=1.05cm},
  responsebox/.style={draw=red!70!black,rounded corners=2pt,fill=red!7,
                  align=center,text width=2.75cm,minimum height=1.38cm},
  ampbox/.style={draw=violet!75!black,dashed,rounded corners=2pt,
                fill=violet!6,align=center,text width=3.35cm,
                minimum height=1.15cm}]
  \node[flowbox] (ain) at (1.55,1.25)
    {incoming radiative\\[-1pt]partial waves $a_\infty^{\rm in}$};
  \node[propbox] (cin) at (5.25,1.25)
    {exact exterior map\\[-1pt]CHE/NS connection and\\[-1pt]Maxwell reconstruction};
  \node[flowbox] (trace) at (8.95,1.25)
    {disk/rim trace space\\[-1pt]$\mathscr V_{\rm tr}$};
  \node[responsebox] (response) at (8.95,-1.20)
    {retarded disk/rim response\\[-1pt]
     $a_D^{\rm out}=\mathsf R_D a_D^{\rm in}$\\[-1pt]
     \scriptsize microscopic law required};
  \node[propbox] (cout) at (5.25,-1.20)
    {exact exterior map\\[-1pt]CHE/NS connection and\\[-1pt]Maxwell reconstruction};
  \node[flowbox] (aout) at (1.55,-1.20)
    {outgoing radiative\\[-1pt]partial waves and $\mathsf S_{\mathsf R_D}$};
  \node[ampbox] (amp) at (5.25,-3.35)
    {all-spin root--Kerr\\[-1pt]Compton kernel $\cK_{\rm C}$\\[-1pt]
     \scriptsize fixes $\mathfrak R_{\rm min}$ only on the radiative\\[-1pt]
     \scriptsize on-shell trace subspace};

  \draw[->,black!75,line width=.65pt] (ain) -- (cin);
  \draw[->,black!75,line width=.65pt] (cin) -- (trace);
  \draw[->,black!75,line width=.65pt] (trace) -- (response);
  \draw[->,black!75,line width=.65pt] (response) -- (cout);
  \draw[->,black!75,line width=.65pt] (cout) -- (aout);
  \draw[->,violet!75!black,dashed,line width=.65pt] (amp.north east) -- (response.south west);
\end{tikzpicture}
\caption{Boundary-data flow for root--Kerr photon scattering.  Exact
exterior maps propagate radiative waves to and from disk/rim traces, but do
not select a scattering state.  The on-shell root--Kerr Compton amplitude
constrains the response only on the radiative trace subspace.}
\label{fig:boundary-flow}
\end{figure}

The logical separation between exact exterior propagation, source response,
and on-shell amplitude input is summarized in figure~\ref{fig:boundary-flow}.

For a Kerr black hole the analogous notation hides substantially less
physics.  The future event horizon is a null characteristic surface, and
regularity in ingoing Kerr coordinates selects the purely ingoing branch with
corotating frequency $\omega-m\Omega_H$.  In the convention of
\eqref{eq:general-disk-reflection}, this is the horizon prescription
$\mathsf R_H=0$.  It is therefore a geometrically selected physical
condition rather than a freely adjustable material parameter.

No corresponding step is available in root--Kerr.  The physical inner locus
is the ordinary ODE point $r=0$, interpreted as two faces of a material disk
and supplemented by a distributional rim.  It is neither a horizon nor a
one-way causal boundary.  Consequently the disk-side labels in
\eqref{eq:diskconnectionblocks} are a useful parameterization, but they are
not selected by local regularity of the exterior equation; analytic
continuation through $r=0$ is likewise not a scattering condition.  The
nontrivial task in the root--Kerr problem is precisely to determine or match
$\mathsf R_D$.  Once it is supplied by a regulated microscopic model, or by
a specified on-shell completion on the radiative trace subspace,
\eqref{eq:response-dressed-S} converts the exact CHE/NS connection data into
phase shifts, helicity conversion, and absorption observables.

\section{Discussion and outlook}
\label{sec:discussion}

The root--Kerr scalar and electromagnetic problems admit a common analytic separation
into exterior propagation and source data, but the two probes use that
separation differently.  The massless charged scalar has the direct bulk
coupling $qQ$ displayed in \eqref{eq:scalarradial}; it is consequently a
benchmark for the CHE/NS connection problem.  The Maxwell
exterior has the same confluent-Heun structure but no bulk coupling to $Q$.
Its charge-dependent scattering is therefore a source problem from the
outset, not a property of the homogeneous radial equation.

The physical source modifies the usual black-hole boundary-value problem in an essential way.  The
radial singular points $r=\pm\ii a$ organize the CHE equation, whereas the
physical object is the two-faced disk together with the associated
distributional edge contribution at its rim.  There
is no horizon and hence no analogue of a universally selected ingoing
solution.  The relevant additional datum is a causal disk/rim response law.
The junction conditions determine the form such a law must take, but not its
susceptibility.

The stationary field gives a further nontrivial constraint on any proposed
completion.  The sharp-regulator calculation in
\eqref{eq:one-ring-superluminal} shows that one subluminal corotating rim
charge cannot retain both $Q$ and the magnetic moment $Qa$.  At least one
independent magnetization/current channel is required.  This result explains
why replacing the magic source by a featureless perfect conductor or a
one-component membrane is not a controlled approximation.

The amplitude information has an equally precise but more limited role.  In
the regulated channel space it imposes the matching condition
\eqref{eq:minimalreaction} on the radiation-visible disk/rim traces.  It does
not determine the response on unprobed trace directions, a rim spectrum, a
damping rate, or a static polarizability.  Consequently this work does not
quote an electromagnetic-wave absorption coefficient, a unique phase shift, or an intrinsic
static or dynamical disk/rim susceptibility for the magic object.

The corresponding single-copy programme is now sharply formulated.  A
regulated disk/rim response law would turn the root--Kerr Compton data and the
connection matrices derived here into conservative, spectral, and absorptive
observables.  It would thereby provide a finite-size completion in the
gauge-theory problem that can subsequently be compared with a proposed
double-copy completion of the Kerr observables.  Such a comparison requires a
matching of the relevant response operators; it is not supplied by the
three-point double-copy relation alone.

A microscopic completion can be tested in a definite sequence.  One first
specifies a finite-speed regulated disk and a two-channel rim action, computes
its retarded trace kernel, and verifies the matching condition
\eqref{eq:minimalreaction}.  Only after that test has passed can the model be
used in \eqref{eq:response-dressed-S} to calculate static response,
resonances, or absorption.  This order of operations keeps exact exterior
propagation, amplitude matching, and material dynamics logically distinct.

\acknowledgments
The author is grateful to Yilber Fabian Bautista, Alok Laddha, and Akavoor
Manu for useful discussions.  OpenAI Codex was used as an assistive tool for
presentation, algebraic cross-checks, and internal-consistency checks during
the preparation of this manuscript.  The author acknowledges support
from the S\~ao Paulo Research Foundation (FAPESP) under Grant No.~2025/01291-6.

\appendix

\section{Separation, NS connection data, and angular projections}
\label{app:nsangular}

\paragraph{Direct scalar separation and NS coefficient matching.}
\label{app:separationdictionary}

The scalar separation in section~\ref{sec:scalar} follows directly from a
short operator calculation.  For the metric \eqref{eq:metric},
\begin{equation}
 \sqrt{-g}=\Sig\sin\theta,\qquad
 \bar A^t=\frac{Qr}{\Sig},\qquad
 \bar A^\phi=\frac{Qar}{\Sig\Del},\qquad
 \nabla_\mu\bar A^\mu=0 .
 \label{eq:scalaroperatorinputs}
\end{equation}
Writing $D_\mu=\nabla_\mu-\ii q\bar A_\mu$ and using
$\nabla_\mu\bar A^\mu=0$, we have
\begin{equation}
 D_\mu D^\mu=\nabla^2-2\ii q\bar A^\mu\partial_\mu
 -q^2\bar A_\mu\bar A^\mu .
 \label{eq:scalaroperatoridentity}
\end{equation}
After inserting \eqref{eq:scalaransatz} and multiplying the result by
$\Sig$, the scalar equation becomes
\begin{align}
0={}&\left\{\partial_r(\Del\partial_r)
 +\frac1{\sin\theta}\partial_\theta
  \bigl(\sin\theta\partial_\theta\bigr)
 +\omega^2\Sig-\frac{m^2\Sig}{\Del\sin^2\theta}
 -2qQ\omega r\right.\nonumber\\[-2pt]
&\left.\hspace{23mm}
 +\frac{2amqQr+q^2Q^2r^2}{\Del}\right\}
 R_{\ell m}(r)S_{\ell m}(\theta) .
 \label{eq:scalarunseparatedoperator}
\end{align}
The identity
\begin{equation}
 \frac{\Sig}{\Del\sin^2\theta}=\frac1{\sin^2\theta}-\frac{a^2}{\Del}
 \label{eq:scalarseparationidentity}
\end{equation}
separates \eqref{eq:scalarunseparatedoperator}: the terms proportional to
$a^2\omega^2\cos^2\theta-m^2/\sin^2\theta$ form the angular equation,
whereas the remaining radial numerator is
$(am+qQr)^2$.  This yields \eqref{eq:scalarangular} and
\eqref{eq:scalarradialdirect} without a Newman--Janis prescription for the
wave equation.

For both probe sectors, the NS dictionary follows from a coefficient match
after the substitution $z=(r+\ii a)/(2\ii a)$.  The finite-pole exponents
fix the two monodromies, the power of the two asymptotic waves fixes $m_3$,
and the remaining regular coefficient fixes the accessory parameter $u$.
In the $N_f=3$ convention used in \eqref{eq:connectioncoefficient}, the
comparison is
\begin{center}
\begin{tabular}{lcc}
\toprule
 & massless charged scalar & Maxwell perturbation ($s=\pm1$) \\
\midrule
$a_{{\rm NS},0}$ & $(m-\ii qQ)/2$ & $(m+s)/2$ \\
$a_{{\rm NS},1}$ & $(m+\ii qQ)/2$ & $(m-s)/2$ \\
$m_3$ & $-\ii qQ$ & $-s$ \\
$u$ & $-\lambda_0+q^2Q^2-L(m+\ii qQ)/2$
& $-\lambda_s-s(s+1)-L(m+s)/2$
\\
\bottomrule
\end{tabular}
\end{center}
The first two rows reproduce the indicial roots
\eqref{eq:scalarindices} and \eqref{eq:localexponents}; the third row
reproduces the Coulomb powers at infinity in the scalar problem and the
spin-weighted asymptotic powers in the electromagnetic problem.  The last row is then
fixed by direct comparison of the coefficient with no finite-pole residue.
This establishes \eqref{eq:scalarNSdictionary} and \eqref{eq:dictionary}
in one common convention.

\paragraph{NP decoupling of the extreme Maxwell scalars.}
\label{app:npdecoupling}

We record the elimination step underlying section~\ref{sec:photon}.  We define
\begin{equation}
 \mathcal A=D-2\rho,\quad
 \mathcal B=\bar\delta+\pi-2\alpha,\quad
 \mathcal C=\delta-2\tau,\quad
 \mathcal D=\Delta+\mu-2\gamma .
 \label{eq:npoperatorabbreviations}
\end{equation}
The two $s=+1$ Maxwell equations are $\mathcal A\phi_1-\mathcal B\phi_0=0$
and $\mathcal C\phi_1-\mathcal D\phi_0=0$.  Applying
$\mathcal C$ and $\mathcal A$, respectively, and subtracting gives
\begin{equation}
 [\mathcal C,\mathcal A]\phi_1+
 (\mathcal A\mathcal D-\mathcal C\mathcal B)\phi_0=0 .
 \label{eq:npeliminationidentity}
\end{equation}
The cancellation of $\phi_1$ can be displayed explicitly.  Away from the
disk/rim source, the principal tetrad has
$\kappa=\sigma=\varepsilon=\Psi_1=\Phi_{01}=0$, and the relevant
commutator and Ricci identity are
\begin{equation}
 [\delta,D]=(\bar\alpha+\beta-\bar\pi)D-\bar\rho\,\delta,
 \qquad
 D\tau-\delta\rho
 =\rho(\bar\pi-\bar\alpha-\beta)+\bar\rho\,\tau .
 \label{eq:npcommutatoridentity}
\end{equation}
Expanding the weighted commutator and then using the two first-order Maxwell
equations gives
\begin{align}
 [\mathcal C,\mathcal A]\phi_1
 &=
 \bigl([\delta,D]+2D\tau-2\delta\rho\bigr)\phi_1\nonumber\\
 &=(\bar\alpha+\beta-\bar\pi)\mathcal B\phi_0
 -\bar\rho\,\mathcal D\phi_0 .
 \label{eq:npweightedcommutator}
\end{align}
All terms proportional to $\phi_1$ cancel in the second line.  Substitution
into \eqref{eq:npeliminationidentity} therefore yields the closed
spin-$+1$ operator
\begin{equation}
 \left[
 (D-2\rho-\bar\rho)\mathcal D
 -(\delta+\bar\pi-\bar\alpha-\beta-2\tau)\mathcal B
 \right]\phi_0=0 .
 \label{eq:npclosedphi0}
\end{equation}
The primed pair of NP Maxwell equations gives the corresponding operator for
$\rho^{-2}\phi_2$.  For compactness, we denote these full decoupled
second-order spacetime operators by $\mathcal T_s$: $\mathcal T_{+1}$ is the
operator on the left-hand side of \eqref{eq:npclosedphi0}, while
$\mathcal T_{-1}$ is its primed counterpart acting on $\rho^{-2}\phi_2$.
Inserting
\eqref{eq:separation}, and collecting the $r$- and $\theta$-dependent
terms, yields
\begin{equation}
 \mathcal T_s\psi_s=
 e^{-\ii\omega t+\ii m\phi}
 \left[{}_sS\,\mathcal R_sR_s+R_s\,\mathcal A_s{}_sS\right]=0,
 \label{eq:npseparationidentity}
\end{equation}
where $\mathcal R_s$ and $\mathcal A_s$ are precisely the radial and angular
operators displayed in \eqref{eq:radial} and \eqref{eq:angular}.  The
separation constant is fixed by requiring the two brackets in
\eqref{eq:npseparationidentity} to equal $-\lambda_sR_s$ and
$+\lambda_s{}_sS$, respectively.  This derivation also explains why the
$s=-1$ field must be $\rho^{-2}\phi_2$: without that rescaling the primed
operator does not take the common Teukolsky form.

\paragraph{Angular perturbation theory and finite angular-momentum mixing matrices.}
\label{app:angular}

The angular equation \eqref{eq:angular} can be handled without numerical
spin-weighted spheroidal functions at low order.  In the spin-weighted
spherical basis, we define
\begin{equation}
 A^{(s,m)}_\ell\equiv
 \frac{\sqrt{(\ell^2-m^2)(\ell^2-s^2)}}
 {\ell\sqrt{(2\ell-1)(2\ell+1)}},
 \qquad
 d^{(s,m)}_\ell\equiv-\frac{ms}{\ell(\ell+1)} .
 \label{eq:wignercoeffs}
\end{equation}
Then multiplication by $\cos\theta$ is represented by the tridiagonal
matrix
\begin{equation}
 [X^{(s,m)}]_{\ell'\ell}=
 A^{(s,m)}_\ell\delta_{\ell',\ell-1}
 +d^{(s,m)}_\ell\delta_{\ell',\ell}
 +A^{(s,m)}_{\ell+1}\delta_{\ell',\ell+1} .
 \label{eq:cosinematrix}
\end{equation}
The finite bandwidth in \eqref{eq:cosinematrix} is the origin of the
selection rule in \eqref{eq:ladderform}.  It also gives a direct algebraic
check of the angular eigenvalues.  With $L=4a\omega$, the $m=s=1$ radial
separation constants begin as
\begin{align}
 \lambda_{111}&=-\frac{3L}{4}+\frac{9L^2}{320}
 -\frac{3L^3}{2560}+O(L^4),\nonumber\\
 \lambda_{211}&=4-\frac{7L}{12}+\frac{479L^2}{12096}
 +\frac{611L^3}{870912}+O(L^4),\nonumber\\
 \lambda_{311}&=10-\frac{11L}{24}+O(L^2).
 \label{eq:angularchecks}
\end{align}
The first two entries agree with \eqref{eq:lambda111}--\eqref{eq:lambda211}.
The distinction between the spheroidal angular eigenvalue ${}_sA_{\ell m}$
and the radial constant $\lambda_s$ in \eqref{eq:defKlambda} must be kept
throughout this expansion.

For the local flip sector, we define two commuting linear operations on
partial-wave matrices, with $X_{\rm out}=X_{-1}^m$ and
$X_{\rm in}=X_{+1}^m$,
\begin{equation}
 \mathscr X[A]=(a\omega X_{\rm out})A+A(a\omega X_{\rm in}),\qquad
 \mathscr Y[A]=(a\omega X_{\rm out})A-A(a\omega X_{\rm in}).
 \label{eq:superoperators}
\end{equation}
They commute.  The entire function in \eqref{eq:localamplitude} therefore
defines $E(\mathscr X,\mathscr Y,2a\omega)$ unambiguously by its convergent
power series.  Equivalently, every homogeneous polynomial $E_n(x,y,z)$ is
applied to $\cU_{\rm loc}$ by the replacement
$(x,y,z)\mapsto(\mathscr X,\mathscr Y,2a\omega)$.  This gives the exact
finite-partial-wave implementation of the all-spin contact kernel; the
truncation in \eqref{eq:lift} is used only when a fixed spin order is
required, and needs no angular quadrature.

\paragraph{Entire local factor and the first Wigner bands.}
\label{app:allspinlift}

For the pole prescription of section~\ref{sec:response}, the entire function
in \eqref{eq:localamplitude} is
\begin{equation}
 E(x,y,z)=\sum_{\varepsilon=\pm1}
 \frac{e^{\varepsilon y}-e^x\cosh z+
 (x-\varepsilon y)e^x\operatorname{sinhc}z}
 {(x-\varepsilon y)^2-z^2} .
 \label{eq:exactentirefactor}
\end{equation}
Although the individual fractions in \eqref{eq:exactentirefactor} are not
regular term by term, their sum is entire.  A coefficient definition that
makes this regularity manifest is
\begin{equation}
 E_n(x,y,z)=\sum_{\varepsilon=\pm1}
 \frac{[t^{n+2}]\left[e^{t\varepsilon y}-e^{tx}\cosh(tz)
 +(x-\varepsilon y)e^{tx}\frac{\sinh(tz)}{z}\right]}
 {(x-\varepsilon y)^2-z^2},
 \qquad E=\sum_{n\geq0}E_n .
 \label{eq:entirefactorcoefficients}
\end{equation}
Here $[t^p]$ denotes the coefficient of $t^p$.  Under the simultaneous
scaling $(x,y,z)\mapsto t(x,y,z)$, the denominator carries a common $t^2$;
this is why the coefficient extraction begins at $t^{n+2}$.  In particular,
\begin{equation}
 E_0=1,\qquad E_1=\frac23x,\qquad
 E_2=\frac1{12}\left(3x^2+y^2+z^2\right).
 \label{eq:firstentirecoefficients}
\end{equation}
Equations \eqref{eq:entirefactorcoefficients} and
\eqref{eq:superoperators} are a direct finite algebraic prescription for
every coefficient in \eqref{eq:lift}.

The first two entries of \eqref{eq:ladder} provide useful transparent checks.
For the $(\ell',\ell,m)=(1,1,1)$ block, $E_0=1$ gives
\begin{equation}
 [\cK_{\rm loc}^{m=1}]_{1-,1+}
 =-\frac{2a^2\omega^3}{3\pi M}=a^2c_1,
 \qquad c_1=-\frac{2\omega^3}{3\pi M} .
 \label{eq:firstladdercoefficient}
\end{equation}
For the transition $(\ell',\ell,m)=(2,1,1)$, only the left action in
$\mathscr X$ can change the outgoing angular momentum of the seed.  Since
\begin{equation}
 A^{(-1,1)}_2=\frac{3}{2\sqrt{15}},
 \label{eq:firstwignerentry}
\end{equation}
the $E_1$ term gives
\begin{align}
 [\cK_{\rm loc}^{m=1}]_{2-,1+}^{(a^3)}
 &=\frac23(a\omega)A^{(-1,1)}_2
 \left(-\frac{2a^2\omega^3}{3\pi M}\right)\nonumber\\
 &=-\frac{2a^3\omega^4}{3\sqrt{15}\pi M}
 =a^3c_2 .
 \label{eq:secondladdercoefficient}
\end{align}
This proves the displayed values of $c_1$ and $c_2$ without further angular
integration.  The higher $c_i$ follow from the same coefficient prescription
and the tridiagonal multiplication matrix \eqref{eq:cosinematrix}.

\paragraph{Projection and normalization of the quadrupole flip seed.}
\label{app:quadrupoleseed}

The normalization in \eqref{eq:allspinlift} follows directly from the
unit-flux projection \eqref{eq:localprojection}.  The leading local
amplitude is obtained from \eqref{eq:localamplitude} by setting $E=E_0=1$:
\begin{equation}
 \mathcal M_{{\rm loc},(2)}^{-+}
 =Q^2\left(A^2-\frac{C^2z^2}{P^2}\right).
 \label{eq:quadrupolelocalamplitude}
\end{equation}
We choose the spin axis to be $a\hat{\boldsymbol z}$.  Each factor of
$\chi^\mu$ carries one spin-one Wigner matrix on the incoming sphere and
one on the outgoing sphere.  We use the standard relation
\begin{equation}
 {}_sY_{\ell m}(\Omega)=(-1)^s
 \sqrt{\frac{2\ell+1}{4\pi}}\,D^\ell_{m,-s}(\Omega)
 \label{eq:spinweightedWigner}
\end{equation}
and the product decomposition of two $D^1$ matrices.  The scalar and
rank-two components cancel in the opposite-helicity contraction in
\eqref{eq:quadrupolelocalamplitude}.  The remaining rank-one contribution has
support only at $\ell'=\ell=1$; it vanishes for $m=0$, while axial symmetry
and parity equate the $m=\pm1$ coefficients.  The Wigner decomposition is
therefore
\begin{equation}
 \mathcal M_{{\rm loc},(2)}^{-+}(\Omega',\Omega)
 =\mathcal N Q^2a^2\omega^2
 \sum_{n=\pm1}{}_{+1}Y_{1n}(\Omega')\,
 {}_{-1}Y_{1n}^{*}(\Omega).
 \label{eq:quadrupolewignerkernel}
\end{equation}
The coefficient $\mathcal N$ is fixed algebraically.  At the reference point
$\Omega=\Omega'=(\pi/2,0)$ in the opposite-helicity spin frame,
$\chi^0=2\omega$ and $\chi^3=0$, so that
\begin{equation}
 \mathcal M_{{\rm loc},(2)}^{-+}=-4Q^2a^2\omega^2,
 \label{eq:quadrupolecalibration}
\end{equation}
At this point the needed reduced Wigner elements are
\begin{equation}
 d^1_{1,-1}\left(\frac\pi2\right)
 =d^1_{-1,-1}\left(\frac\pi2\right)
 =d^1_{1,1}\left(\frac\pi2\right)
 =d^1_{-1,1}\left(\frac\pi2\right)=\frac12.
 \label{eq:quadrupole-d-elements}
\end{equation}
Equation \eqref{eq:spinweightedWigner} then gives
\begin{equation}
 \sum_{n=\pm1}{}_{+1}Y_{1n}(\Omega')\,
 {}_{-1}Y_{1n}^{*}(\Omega)
 =2\left(\sqrt{\frac{3}{4\pi}}\frac12\right)^2
 =\frac{3}{8\pi}.
 \label{eq:quadrupole-harmonic-calibration}
\end{equation}
Comparison with \eqref{eq:quadrupolewignerkernel} fixes
$\mathcal N=-32\pi/3$.  This establishes the Wigner coefficient without
numerical angular integration.

We may now evaluate the projection itself.  For $\lambda=+$ and
$\lambda'=-$, substitution of \eqref{eq:quadrupolewignerkernel} into
\eqref{eq:localprojection} gives
\begin{align}
 [\cU_{\rm loc}^{m}]_{1-,1+}
 &=\frac{\omega}{16\pi^2MQ^2}
 \left(-\frac{32\pi}{3}Q^2a^2\omega^2\right)
 \sum_{n=\pm1}
 \left[\int\dd\Omega'\,{}_{+1}Y_{1m}^{*}
 {}_{+1}Y_{1n}\right]
 \left[\int\dd\Omega\,{}_{-1}Y_{1n}^{*}
 {}_{-1}Y_{1m}\right]\nonumber\\
 &= -\frac{2a^2\omega^3}{3\pi M}\,
 \begin{cases}
  1,&m=\pm1,\\
  0,&m=0,
 \end{cases}
 \label{eq:quadrupoleseedprojection}
\end{align}
where orthonormality of the spin-weighted harmonics was used in the second
line.  The $m=1$ component is precisely \eqref{eq:allspinlift}.  This
derivation also fixes the common external-photon phase convention used in the
opposite-helicity block.

\paragraph{Integer-\texorpdfstring{$\ell$}{ell} NS regularization.}
\label{app:nsregularization}

At physical integer angular momentum, the local exponents differ by integers,
and the second Frobenius solution can contain a logarithmic contribution.  A
direct substitution of integer $\ell$ into the gamma functions of
\eqref{eq:connectioncoefficient} creates spurious pole-over-pole forms.  The
regulated procedure is algebraic.  For every gamma function approaching a
nonpositive integer, we first use
\begin{equation}
 \Gamma(-n+x)=\frac{\Gamma(1+x)}{x\prod_{j=1}^{n}(x-j)},
 \qquad n\in\mathbb Z_{\geq0},
 \label{eq:gammapole}
\end{equation}
We write every noninteger power as
\begin{equation}
 L^{p(L)}=L^{p(0)}\exp\{[p(L)-p(0)]\log L\}.
 \label{eq:powerregularization}
\end{equation}
Only then is the low-$L$ expansion taken.  The result is a canonical
regular/logarithmic local basis and a finite $2\times2$ connection matrix at
each physical $\ell$.  In the $(s,m)=(1,1)$ sector its regular-row behavior
is
\begin{equation}
 C_{\rm reg}^{(\ell)}\sim L^{-\ell-1/2},\qquad \ell=1,\ldots,7,
 \label{eq:regscaling}
\end{equation}
while the second row contains the corresponding logarithmic terms.  This
construction is independent of a horizon boundary condition and is therefore
appropriate for the root--Kerr exterior.

\section{Disk/rim response kernels and their channel projection}
\label{app:diskresponse}

\paragraph{Regulated stationary disk and rim constraint.}
\label{app:rimregulator}

We give the intermediate steps leading to
\eqref{eq:cut-disk-charge}--\eqref{eq:rim-magnetization-channel}.  They
follow from the stationary source alone and do not use a perturbative response
ansatz.  With $u=a^2-b^2$, the regulated disk charge obtained from
\eqref{eq:magic-source} is
\begin{align}
 Q_D(b)
 &=2\pi\int_0^b R\,\dd R\,\sigma_D(R)
 =-Qa\int_0^b\frac{R\,\dd R}{(a^2-R^2)^{3/2}}\nonumber\\
 &=-Q\left(\frac{a}{\sqrt u}-1\right).
 \label{eq:app-cut-disk-charge}
\end{align}
Using $J_\phi=(R/a)\sigma_D$, its magnetic moment is
\begin{align}
 \mu_D(b)
 &=\pi\int_0^bR^2\,\dd R\,J_\phi(R)
 =-\frac Q2\int_0^b\frac{R^3\,\dd R}{(a^2-R^2)^{3/2}}\nonumber\\
 &=-\frac Q2\left(\frac{a^2}{\sqrt u}+\sqrt u-2a\right).
 \label{eq:app-cut-disk-moment}
\end{align}
For a ring at $R=b$ with charge $q_{\rm r}$ and azimuthal current
$I_{\rm r}$, charge and magnetic-moment matching require
\begin{equation}
 Q_D+q_{\rm r}=Q,
 \qquad \mu_D+\pi b^2I_{\rm r}=Qa .
 \label{eq:app-ring-matching-equations}
\end{equation}
Substituting \eqref{eq:app-cut-disk-charge} and
\eqref{eq:app-cut-disk-moment} gives \eqref{eq:one-ring-match}.  The
resulting charge-carrier speed is
\begin{align}
 v_{\rm r}
 &=\frac{2\pi bI_{\rm r}}{q_{\rm r}}
 =\frac{a^2+u}{ab}
 =\frac{a^2+u}{a\sqrt{a^2-u}},
 \label{eq:app-ring-speed}
\end{align}
which is strictly greater than one for $0<u<a^2$, since
$(a^2+u)^2-a^2(a^2-u)=u(3a^2+u)>0$.  A subluminal corotating charge component
has instead $I_{\rm cor}=\Omega_Dq_{\rm r}/(2\pi)$, and its shortfall is
\begin{align}
 I_{\rm mag}&=I_{\rm r}-I_{\rm cor}
 =\frac{Q\sqrt u}{\pi b^2},
 &\mu_{\rm mag}&=\pi b^2I_{\rm mag}=Q\sqrt u .
 \label{eq:app-rim-magnetization}
\end{align}
This establishes the two-channel conclusion stated in the main text.

\paragraph{Projection of a disk kernel onto photon channels.}
\label{app:traceprojection}

Here we spell out the map which takes a response defined on the physical disk
to the channel-space matrix used in the text.  This is the required bridge in
any attempt to identify an exterior radial connection coefficient directly
with a photon phase shift.  At fixed $(\omega,m)$, let
$\mathsf T^{\rm in/out}_{\ell\lambda m}(R)$ denote the collection of
gauge-invariant Maxwell traces of a unit-flux mode on the two faces of the
disk.  It includes tangential electric and magnetic fields, the appropriate
normal components, and a face label.  Its detailed form follows from the
Newman--Penrose reconstruction in section~\ref{sec:photon} together with
the ordinary-point propagation described in section~\ref{sec:diskmatching}.

\paragraph{Maxwell reconstruction entering the trace map.}
\label{app:maxwellreconstruction}

The trace map is defined for a complete Maxwell field, rather than for one
extreme NP scalar in isolation.  With the tetrad conventions of
\eqref{eq:tetrad}, the radiative data obey the electromagnetic
Teukolsky--Starobinsky identities
\cite{Starobinskil:1974nkd,Teukolsky:1973ha}
\begin{align}
 \Del(D_0^\dagger)^2\Del R_{+1}&=\mathcal C_{\ell m\omega}^*R_{-1},
 & (D_0)^2R_{-1}&=\mathcal C_{\ell m\omega}R_{+1},\nonumber\\
 D_0&=\partial_r-\ii\frac{K}{\Del},
 & D_0^\dagger&=\partial_r+\ii\frac{K}{\Del} .
\label{eq:rootkerrTS}
\end{align}
Here $K(r)=\omega(r^2+a^2)-am$ and $\Del=r^2+a^2$, as in
\eqref{eq:defKlambda}; $K(r)$ is the separated radial kinematic
combination and must not be confused with the partial-wave kernels
$\cK$.
These are the separated form of the first-order NP Maxwell reconstruction
relations: the second-order radial intertwiner $(D_0)^2$ maps an $s=-1$
solution to its $s=+1$ partner, while the adjoint intertwiner
$\Del(D_0^\dagger)^2\Del$ maps back.  The separation constant fixes their
relative normalization through $\mathcal C_{\ell m\omega}$.
The relative normalization of $R_{+1}$ and $R_{-1}$ is therefore fixed for
one radiative Maxwell solution.  The middle NP scalar is then fixed, up to the
non-radiative Coulomb zero mode, by the first-order NP equations.  The
resulting real field is reconstructed as
\begin{equation}
 f_{\mu\nu}=2\left[
 \phi_1\bigl(n_{[\mu}\ell_{\nu]}+m_{[\mu}\bar m_{\nu]}\bigr)
 +\phi_2\ell_{[\mu}m_{\nu]}
 +\phi_0\bar m_{[\mu}n_{\nu]}
 \right]+\mathrm{c.c.}
 \label{eq:maxwellNPReconstruction}
\end{equation}
The phase of the nonzero constant $\mathcal C_{\ell m\omega}$ is a tetrad
convention.  It cancels from physical traces provided the same reconstruction
convention is used for both asymptotic and disk-side bases.

This propagation admits a compact disk-first construction.  We take
$\Psi_A$, $A={\rm reg},{\rm log}$, to be the direct raw local CHE basis.  We
define the upper-face boundary phase-space trace matrix by
\begin{equation}
 H_{\alpha A}(\theta)=
 \left.
 \left\{f_{\theta t}+\Omega_D f_{\theta\phi},\,
 f_{\phi t},\,n_\mu f^{\mu\theta},\,n_\mu f^{\mu\phi}
 \right\}_{\!\alpha}[\Psi_A]
 \right|_{r=0^+} .
\label{eq:rawhertztrace}
\end{equation}
The row index $\alpha=1,\ldots,4$ labels the displayed trace components,
and the column label $A={\rm reg},{\rm log}$ labels the two local CHE
solutions.  On the disk $r=0$, the physical cylindrical radius is
$R=a\sin\theta$ and the corotating angular velocity is
$\Omega_D=1/a$.
The first two rows are the corotating tangential-electric-field components;
the remaining two are the normal electric-flux traces that enter the Maxwell
jump condition.
If $C_{A\sigma}$ is the NS connection matrix from this raw local basis to the infinity in/out basis
$\sigma={\rm in,out}$, the corresponding upper-face trace map is
\begin{equation}
 \mathsf T^{(+)}_{\alpha\sigma}(\theta)=
 \bigl[H(\theta)\,(C^T)^{-1}\bigr]_{\alpha\sigma}.
 \label{eq:ns-hertz-trace-map}
\end{equation}
All entries in \eqref{eq:ns-hertz-trace-map} are finite after the
integer-$\ell$ gamma-pole resolution: $H$ depends only on a finite set of
radial values and derivatives at the ordinary point, together with the
angular function related by the Teukolsky--Starobinsky identity
\cite{StarobinskyChurilov:1974,Teukolsky:1973ha}.  The
lower-face trace follows by $\theta\mapsto\pi-\theta$ with the fixed normal
orientation.  No disk boundary condition is used in this construction.

The retarded response kernel acts on this finite set of face and tensor
indices, but is in general bilocal in disk radius:
\begin{equation}
 \delta j_a(R)=Q^2\int_0^a R'\dd R'\,
 \Pi^{ab}_{D+R}(R,R';\omega,m)\,f_b(R') .
 \label{eq:radialkernel}
\end{equation}
Here $R$ is the physical cylindrical disk radius.  The lower-case indices
$a,b$ label trace/current components and are unrelated to the spin parameter
$a$.  The quantity $f_b(R')$ is the corresponding incident trace vector.
The measure $R'\dd R'$ is the axisymmetric disk measure after the azimuthal
Fourier mode has been fixed.
Its projection onto radiative channels is consequently
\begin{align}
 [\Pi^m_{\rm ch}]_{\ell'\lambda',\ell\lambda}
 ={}&\int_0^a R\dd R\int_0^a R'\dd R'\,
 \bigl[\mathsf T^{\rm out}_{\ell'\lambda'm}(R)\bigr]^\dagger
 \Pi_{D+R}(R,R';\omega,m)
 \mathsf T^{\rm in}_{\ell\lambda m}(R') .
 \label{eq:traceprojection}
\end{align}
The overall normalization in the conversion from $\Pi^m_{\rm ch}$ to the
order-$Q^2$ channel kernel $\cK^{(2),m}$ is fixed once a unit-flux convention and the Maxwell jump
convention \eqref{eq:kernelbc} are fixed.  Crucially, this
projection preserves $(\omega,m)$ but not $\ell$ or helicity.  The ordinary
point at $r=0$ therefore generates a matrix boundary problem even though
each separated exterior radial equation is second order.

The two-sided structure is made explicit by arranging each incoming or
outgoing trace map into a block vector,
\begin{equation}
 \mathsf T^{A}_{\ell\lambda m}=
\begin{pmatrix}
  \mathsf T^{A,(+)}_{\ell\lambda m}\\[1mm]
  \mathsf T^{A,(-)}_{\ell\lambda m}
\end{pmatrix},\qquad
 A={\rm in,out},\qquad
 \Pi_{D+R}=
 \begin{pmatrix}
  \Pi_{++}&\Pi_{+-}\\
  \Pi_{-+}&\Pi_{--}
 \end{pmatrix}.
 \label{eq:facedecomposition}
\end{equation}
The labels $\pm$ in \eqref{eq:facedecomposition} refer to the upper and
lower \emph{faces} of the disk, not photon helicity.  Local parity, reality,
and retardedness impose relations between these four entries.  They do not,
however, set the off-diagonal face blocks to zero: a rim mode can transmit
stress and charge between the two faces.  This is precisely where a simple
``reflect at $r=0$'' prescription loses physical information.

For a regulated local model the kernel may be expanded in radial derivatives,
for example
\begin{equation}
 \Pi_{D+R}^{ab}(R,R')=
 \pi^{ab}_0(R)\frac{\delta(R-R')}{R}
 +\pi^{ab}_2(R)\frac{D_R^2\delta(R-R')}{R}+\cdots
 +\Pi^{ab}_{\rm rim}(R,R') .
\label{eq:derivativekernel}
\end{equation}
In this illustrative derivative expansion, $D_R$ denotes the radial
derivative acting on the relevant disk tensor component; for a scalar
component it is simply $\partial_R$.  The factor $1/R$ ensures that the
delta function acts as the identity with the radial measure $R\dd R$.
The final term has support at $R=R'=a$ and includes a possible boundary
layer.  Substitution of \eqref{eq:derivativekernel} into
\eqref{eq:traceprojection} gives a finite channel matrix after radial
integration by parts.  Conversely, the on-shell amplitude matching fixes only the
on-shell values of these integrated combinations.  It cannot determine the
individual profiles $\pi_n(R)$ or the off-shell rim kernel, which is the
precise sense in which the effective response is underdetermined
microscopically.

\paragraph{Boundary-to-scattering algebra.}
\label{app:boundarySderivation}

The response-dressed scattering formula
\eqref{eq:response-dressed-S} follows by elementary block elimination.  Let
$a_D^{\rm in/out}$ be the coefficients in the two disk-side bases of
\eqref{eq:diskconnectionblocks}.  The continuation to infinity gives
\begin{align}
 a_\infty^{\rm in}&=\mathsf C_{{\rm in},{\rm in}}a_D^{\rm in}
 +\mathsf C_{{\rm in},{\rm out}}a_D^{\rm out},\nonumber\\
 a_\infty^{\rm out}&=\mathsf C_{{\rm out},{\rm in}}a_D^{\rm in}
 +\mathsf C_{{\rm out},{\rm out}}a_D^{\rm out} .
 \label{eq:boundaryblockcontinuation}
\end{align}
Using $a_D^{\rm out}=\mathsf R_Da_D^{\rm in}$, the first line determines
\begin{equation}
 a_D^{\rm in}=
 \bigl(\mathsf C_{{\rm in},{\rm in}}
 +\mathsf C_{{\rm in},{\rm out}}\mathsf R_D\bigr)^{-1}
 a_\infty^{\rm in} .
 \label{eq:diskcoefficientelimination}
\end{equation}
Substitution into the second line of \eqref{eq:boundaryblockcontinuation}
is exactly \eqref{eq:response-dressed-S}.  This derivation makes no
assumption that $\mathsf R_D$ is diagonal in disk face, photon helicity, or
angular momentum.

\bibliographystyle{JHEP}
\bibliography{subnsoft}

\end{document}